\documentclass[fleqn,usenatbib]{mnras}

\usepackage[T1]{fontenc}
\usepackage{ae,aecompl}

\usepackage{graphicx}	
\usepackage{amsmath}	
\usepackage{amssymb}	
\usepackage{subfigure}
\usepackage{booktabs}

\usepackage[usenames]{xcolor}
\usepackage[normalem]{ulem}

\title[The dynamics of GCs]{Dynamics in the outskirts of four Milky Way globular clusters: \\ 
it's the tides that dominate}

\author[Wan et al.]{Zhen Wan$^{1,2}$\thanks{E-mail: zhen\_wan@ustc.edu.cn},
Anthony~D.~Arnold$^{3}$,
William~H.~Oliver$^{2}$,
Geraint~F.~Lewis$^{2}$,\newauthor
Holger~Baumgardt$^{3}$,
Mark~Gieles$^{4,5}$,
Vincent~H\'{e}nault-Brunet$^{6}$,
Thomas~de~Boer$^{7,8}$, \newauthor
Eduardo~Balbinot$^{9}$, 
Gary~Da~Costa$^{10}$,
Dougal~Mackey$^{10}$,
Denis~Erkal$^{8}$,
Annette~Ferguson$^{11}$,\newauthor
Pete~Kuzma$^{11}$,
Elena~Pancino$^{12}$,
Jorge~Pe{\~n}arrubia$^{11,13}$,
Nicoletta~Sanna$^{12}$,
Antonio~Sollima$^{14}$,\newauthor
Roeland~P.~van~der~Marel$^{15,16}$,
and Laura~L.~Watkins$^{17}$
\\ \\
$^{1}$School of Astronomy and Space Science, University of Science and Technology of China, Hefei, 230026, China\\
$^{2}$Sydney Institute for Astronomy, School of Physics A28, The University of Sydney, NSW 2006, Australia\\
$^{3}$School of Mathematics and Physics, The University of Queensland, St Lucia, QLD 4072, Australia\\
$^{4}$ICREA, Pg. Llu\'{i}s Companys 23, E08010 Barcelona, Spain\\
$^{5}$Institut de Ci\`{e}ncies del Cosmos (ICCUB), Universitat de Barcelona (IEEC-UB), Mart\'{i} Franqu\`{e}s 1, E08028 Barcelona, Spain \\
$^{6}$Department of Astronomy and Physics, Saint Mary's University, 923 Robie Street, Halifax, NS B3H 3C3, Canada\\
$^{7}$Institute for Astronomy, University of Hawaii, 2680 Woodlawn Drive, Honolulu, HI 96822, USA\\
$^{8}$Department of Physics, University of Surrey, Guildford GU2 7XH, UK\\
$^{9}$Kapteyn Astronomical Institute, University of Groningen, Postbus 800, NL-9700AV Groningen, The Netherlands\\
$^{10}$Research School of Astronomy and Astrophysics, Australian National University, Canberra, ACT 2611, Australia\\
$^{11}$Institute for Astronomy, University of Edinburgh, Royal Observatory, Blackford Hill, Edinburgh EH9 3HJ, UK\\
$^{12}$INAF – Osservatorio Astrofisico di Arcetri, Largo E. Fermi 5, I-50125 Firenze, Italy\\
$^{13}$Centre for Statistics, University of Edinburgh, School of Mathematics, Edinburgh EH9 3FD, UK\\
$^{14}$INAF Osservatorio di Astrofisica e Scienza dello Spazio (OAS), Via Gobetti 93/3, 40129 Bologna, Italy\\
$^{15}$Space Telescope Science Institute, 3700 San Martin Drive, Baltimore, MD 21218, USA\\
$^{16}$Center for Astrophysical Sciences, The William H. Miller III Department of Physics \&\\ Astronomy, Johns Hopkins University, Baltimore, MD 21218, USA\\
$^{17}$AURA for the European Space Agency, ESA Office, Space Telescope Science Institute, 3700 San Martin Drive, Baltimore MD 21218, USA
}

\date{Accepted XXX. Received YYY; in original form ZZZ}

\pubyear{2020}

 \makeatother

\begin{document}
\label{firstpage}
\pagerange{\pageref{firstpage}--\pageref{lastpage}}
\maketitle

\begin{abstract}
We present the results of a spectroscopic survey of the outskirts of 4 globular clusters---NGC~1261, NGC~4590, NGC~1904, and NGC~1851--- covering targets within 1 degree from the cluster centres, with 2dF/AAOmega on the Anglo-Australian Telescope (AAT) and FLAMES on the Very Large Telescope (VLT). We extracted chemo-dynamical information for individual stars, from which we estimated the velocity dispersion profile and the rotation of each cluster. The observations are compared to direct $N$-body simulations and appropriate {\sc limepy}/{\sc spes} models for each cluster to interpret the results. In NGC~1851, the detected internal rotation agrees with existing literature, and NGC~1261 shows some rotation signal beyond the truncation radius, likely coming from the escaped stars. We find that the dispersion profiles for both the observations and the simulations for NGC~1261, NGC~1851, and NGC~1904 do not decrease as the {\sc limepy}/{\sc spes} models predict beyond the truncation radius, where the $N$-body simulations show that escaped stars dominate; the dispersion profile of NGC~4590 follows the predictions of the {\sc limepy}/{\sc spes} models, though the data do not effectively extend beyond the truncation radius. The increasing/flat dispersion profiles in the outskirts of NGC~1261, NGC~1851 and NGC~1904, are reproduced by the simulations. Hence, the increasing/flat dispersion profiles of the clusters in question can be explained by the tidal interaction with the Galaxy without introducing dark matter.
\end{abstract}




\section{Introduction}
Globular clusters (GCs) appear to be simple stellar systems in the Milky Way (MW), densely packed with member stars, where stars are typically all found to display a similar age and iron abundance. However, the structure of GCs turns out to be complex---for example, evolving within the Milky Way potential, cluster members have been revealed beyond the tidal radius of GCs, forming stellar envelopes \citep[e.g.,][]{2014MNRAS.442.3044M,2018MNRAS.473.2881K,2019MNRAS.485.4906D} and tidal tails \citep[e.g.,][and references therein]{2021ApJ...914..123I,2022MNRAS.513.3136Z}. Dynamically, internal rotation and the increasing or flattening of velocity dispersion profiles have been observed within some clusters \citep[e.g.][]{2012A&A...538A..18B,2018MNRAS.473.5591K,2018ApJ...860...50F,2018A&A...616A..12G,2018MNRAS.481.2125B,2018ApJ...861...16L,2018ApJ...865...11L,2019MNRAS.485.1460S,2019MNRAS.489..623V}. As an important component of the MW, the underlying reasons that lead to the observed complex structure can also help us to understand the evolution of the parent galaxy.

In the outskirts of the clusters, the tidal interaction between the clusters and the MW is one of the reasons that lead to the observed structures. The MW potential gradually strips member stars from the cluster. During this process, some stars become potential escapers (PEs), i.e. are located within the Jacobi radius, but have energies above the critical energy needed for their escape  \citep{2000MNRAS.318..753F,2001MNRAS.325.1323B,2017MNRAS.466.3937C,2017MNRAS.468.1453D}. Furthermore, stars that are more dynamically heated can escape from their host clusters and form stellar streams along the orbits \citep[e.g.,][ and references therein]{2021ApJ...914..123I}. In addition to the deformation, the MW also changes the dynamical properties of the clusters. Following crossings of the Galactic disk for example, GCs will incur a larger dispersion than they otherwise would in an isolated environment \citep[e.g. $\omega$ Cen, ][]{2012ApJ...751....6D}.

The formation environment also contributes to the shape of the cluster. Theories suggest that GCs can form within dark matter mini halos \citep{1984ApJ...277..470P,2015ApJ...808L..35T}, which if they exist, will inevitably result in a large mass-to-light ratio. The discovery of stellar envelopes surrounding some GCs agrees with the existence of dark matter---these stars are confined to the GC over a long time period \citep{2012MNRAS.419...14C,2017MNRAS.471L..31P,2018MNRAS.473.2881K}. However, different theories suggest that GCs do not necessarily form within dark matter halos. Instead, they can be a natural outcome of the gas turbulence during intense star formation \citep[e.g.,][]{1997ApJ...480..235E,2005ApJ...623..650K, 2015MNRAS.454.1658K}, without the necessity to introduce dark matter content. Besides, the presence of the tidal tails does not agree with the dark matter scenario---stars being able to escape suggests that the dark matter content of GCs that confine the members within the cluster is limited \citep{1996ApJ...461L..13M}.

Measuring the dynamical properties including rotation and dispersion is a quantitative way to reveal the features of GCs: a spectroscopic survey of GC members at the periphery has the advantage of precisely measuring the dynamics, revealing the presence of potential escapers, and estimating the dispersion and the rotation profiles. With the AAT and the 2dF/AAOmega instrument combination, we have obtained moderate-resolution spectra of red giant branch stars in five globular clusters: NGC~3201, NGC~1261, NGC~4590, NGC~1904, and NGC~1851. In the first paper \citep{2021MNRAS.502.4513W} of this survey, we found that the line-of-sight (LOS) velocity dispersion profile of NGC~3201 is elevated in the outskirts compared to the predictions of ditect $N$-body simulations, whereas the inner part fits the data well. We have tested different scenarios that can contribute to an increased velocity dispersion, and found that the tidal interaction with the MW, the escape rate (like the dynamical effect from black holes), and binaries can all increase the dispersion in the outskirts, but cannot fully explain the observed dispersion in NGC~3201. In this paper, we summarise the properties and results for the other four GCs.

This paper is structured as follows: Sec.~\ref{sec:data} summarises how we obtained and reduced the spectra; Sec.~\ref{sec:simulation} discusses the setup of the simulations we performed for the clusters; Sec.~\ref{sec:results} presents the dynamical properties of each GC, and Sec.~\ref{sec:conclusion} concludes and presents a discussion of the results.

\begin{figure*}
    \centering
    \includegraphics[width=2\columnwidth]{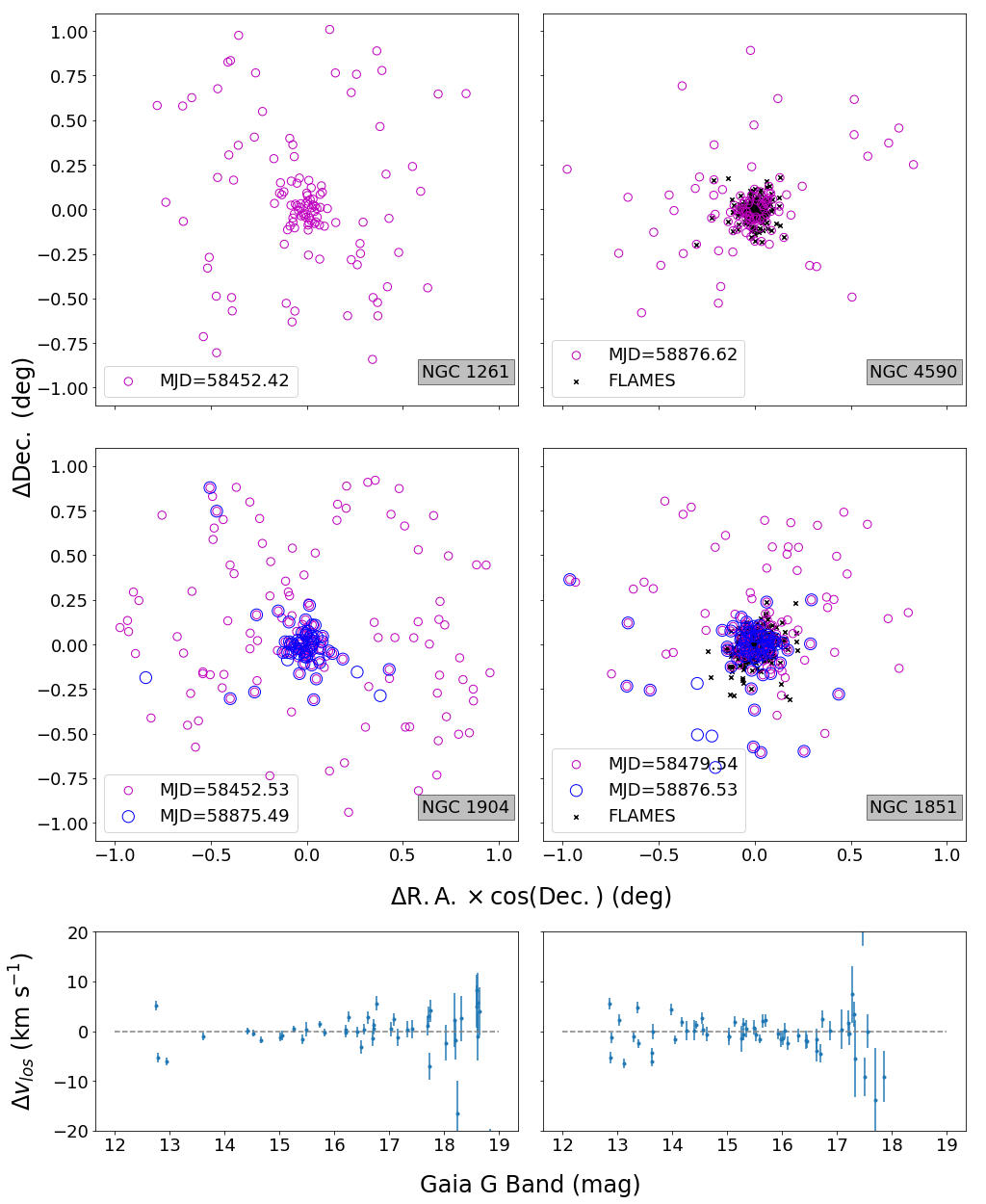}
    \caption{The sky distribution of all the targets observed with 2dF/AAOmega and FLAMES. The different circles indicate the targets at different observational epochs with AAT, which are labelled by the Modified Julian Date (MJD), while the FLAMES sources are shown by black crosses. The bottom two panels show the velocity difference of stars that have been observed twice from NGC~1904 (left) and NGC~1851 (right) and the error bars show the velocity uncertainties. Some candidate binaries can be identified via a large velocity difference. 
    }
    \label{fig:double_epoch}
\end{figure*}

\section{Data}
\label{sec:data}
In this section, we briefly present the data collection and reduction process of the survey. The main data set is from the multi-fibre spectroscopic instrument combination 2dF/AAOmega at the AAT (see \citet{2021MNRAS.502.4513W} for more details). The selection of targets for this spectroscopic follow-up is based on GC members identified by \citet{2019MNRAS.485.4906D} in {\it Gaia} DR2 data. The member stars were identified by the application of a `matched-filter' algorithm to the colour-magnitude diagrams (CMD), which selects members based on the expected location in the CMD from the stellar isochrones with corresponding distance, age and metallicity of each cluster. {\it Gaia} DR2 proper motions are also used to calculate the membership probabilities. We selected red giant branch stars in each GC, so that their spectra will possess the Ca \textsc{ii} triplet (CaT) lines at 8498.02 \hbox{\AA}, 8542.09 \hbox{\AA} and 8662.14 \hbox{\AA} \citep{edlen1956spectrum}, from which we can also estimate the metallicity of the members. 

Tab.~\ref{tab:observation_log} summarises the observations with AAT. The data in question come from the grating 1700D, which provides a spectral resolution of $R \approx 10,000$ and covers the wavelength range from 845 nm to 900 nm. The observations for NGC~1851 and NGC~1904 were split into two epochs---the position of the targets observed in the two epochs for the two GCs, as well as the velocity difference between each epoch, are presented in Fig~\ref{fig:double_epoch}. The detailed target selection, observation, and the raw data reduction with the \textsc{2dfdr}\footnote{\url{https://www.aao.gov.au/science/software/2dfdr}} pipeline \citep{2015ascl.soft05015A}, are described in \citet{2021MNRAS.502.4513W}.

The chemo-dynamical information, including radial velocity, is extracted from the CaT absorption line profiles. For this, we model each spectrum as a combination of the CaT lines and a continuum. The continuum is fitted by means of a 6$^{\rm th}$-order polynomial with major spectral lines being masked out. We then normalise the flux of each spectrum to the best-fitting continuum. Then, we represent each line with a pseudo-Voigt profile---$\mathcal{F}(\lambda)$, the summation of a Gaussian and a Lorentzian profile as an approximation of the Voigt profile, which is often used for calculations of experimental spectral line shapes:

\begin{gather}
    \mathcal{F}(\lambda) = A_0 \mathcal{G}(\lambda,\lambda_0,\sigma_{\rm g}) + A_1 \mathcal{L}(\lambda,\lambda_0,\sigma_{\rm l}) \notag \\
    \mathcal{G}(\lambda,\lambda_0,\sigma_{\rm g}) = \frac{1}{\sqrt{2\pi}\sigma_{\rm g}}\mathrm{e}^{-(\lambda - \lambda_0)^2/(2\sigma_{\rm g}^2)} \notag \\
    \mathcal{L}(\lambda,\lambda_0,\sigma_{\rm l}) = \frac{\sigma_{\rm l}}{\pi((\lambda - \lambda_0)^2 + \sigma_{\rm l}^2)} \notag \\
    \lambda_0 = \lambda_{\textsc{lab}}\times(1 + z)
\end{gather}

where $z$ is the redshift, which is related to the line-of-sight velocity in the low-velocity regime through $z=v/c$.
The $A_{0}$ and $A_{1}$ parameters are the strength of the Gaussian ($\mathcal{G}(\lambda,\lambda_0,\sigma_{\rm g})$) and Lorentzian ($\mathcal{L}(\lambda,\lambda_0,\sigma_{\rm l})$) profiles respectively; $\lambda$ is the wavelength, and $\lambda_0$ is the spectral line centre; $\sigma_{\rm g}$ and $\sigma_{\rm l}$ indicate the line-width from the Gaussian and Lorentzian profiles respectively. The spectral template is constructed with three pseudo-Voigt profiles, whose line centres are correlated by the redshift.

We fit each spectrum with the CaT profile above. The data uncertainties---from the \textit{variance} from \textsc{2dfdr}---are taken into account by convolving with the spectrum profile parameters' probability distribution. The best-fitting line profile parameters and their uncertainties (defined as the mean and the 1$\sigma$ quantiles of the marginalized posterior probability distributions) were derived by 12,000-step MCMC sampling of the posterior using {\sc emcee} \citep[][]{2013PASP..125..306F} (with 2,000 burn-in steps). Here we define a \textit{good\_star} when
\begin{gather}
    S/N > 3\ {\rm and} \notag \\
    \sigma_{\rm vlos} < 3\,\mathrm{km\,s^{-1}}
\end{gather}
where the $S/N$ is the signal-to-noise ratio of the strength of the CaT lines to the residual between the spectrum and the best-fitting model. The $\sigma_{\rm vlos}$ is the LOS velocity uncertainty derived from the spectrum redshift. 

In the lower panel of Fig.~\ref{fig:double_epoch}, we show the velocity difference of the stars that have two epochs of observations. For both NGC~1904 and NGC~1851, the bright stars tend to have a larger velocity difference. The bias towards the brighter side is not likely due to binaries---no evidence exists of a significantly higher binary frequency at the brighter side of the red giant branch in globular clusters \citep[e.g.,][]{2019A&A...632A...3G}. Instead, this might be due to the atmospheric motions in luminous red giant stars including convection and pulsation. For example, the velocity "jitter" at the low surface gravity end of the RGB stars might contribute up to $\geq\ 1 \mathrm{km\ s^{-1}}$ \citep[e.g.,][]{2008A&A...480..215H} to the velocity variance. It is known that some metal-poor red giants exhibit velocity jitter at amplitudes to the level of $\sim1.5-2 \mathrm{km\ s^{-1}}$. This phenomenon seems to affect the intrinsically brightest stars only, within ~1 mag from the tip of the red giant branch \citep[][]{2003AJ....125..293C} and it is unfortunately indistinguishable from RV variations due to tbe effect from binaries, though this effect alone is not enough to explain the velocity difference we observed. This paper will not explore the detail of the bias, instead, to avoid this bias, we only include the members with a Gaia G magnitude $\mathrm{G} > 14\ \mathrm{mag}$ in the dynamical analyses below. In addition, combining stars with two epochs observation, we found a $0.99\ \mathrm{km\ s^{-1}}$ systematic uncertainty of the LOS velocity, which is taken into account in the following analyses of the dispersion and the rotation profiles.

\begin{table*}
    \centering
    \begin{tabular}{c|c|c|c|c|c}
        \hline
        Target Name & Epochs & Exposure Time & $\mathrm{N_{target}}$ & $\mathrm{N_{good star}}$ & MJD  \\
        \hline
        NGC~1261 & 1 & 7800s & 138 & 78 &  58452.42\\
        NGC~1851 & 2 & 7200s/ 1800s & 126/74 & 95/58 & 58479.54/ 58876.53\\
        NGC~1904 & 2 & 7800s/ 7200s & 188/ 77 & 121/ 45 & 58452.53/ 58875.49 \\
        NGC~4590 & 1 & 8400s & 92 & 76 & 58876.62\\
        \hline
    \end{tabular}
    \caption{The observational details of our GC survey with 2dF/AAOmega, demonstrating multi-epoch observations for two of our GC targets.
    Stars were observed at multiple times across different epochs so as to mitigate the effect of binaries. Note that this table only includes the data from AAT 2dF/AAOmega.
    }
    \label{tab:observation_log}
\end{table*}

For NGC 1851 and NGC 4590 we also obtained FLAMES/Giraffe spectroscopic data (ESO proposal ID: 0102.D-0164A, PI: Gieles), including two 1-hour observations per cluster taken 45 days apart from each other, with a spectral resolution of $R \approx 18000$, and a wavelength coverage of $848.231$ to $889.328$ nm. Note that the raw data from AAOmega and FLAMES were reduced independently. After downloading the pipeline-reduced spectra from the ESO Science Data archive, we used the {\tt Skycorr} package \citep[][]{2014A&A...567A..25N} to perform sky subtraction from the individual spectra. We then co-added the spectra of each star using the {\tt IRAF} {\tt scombine} task and determined the radial velocities with the help of the {\tt IRAF} {\tt fxcor} task. For the cross-correlation with the observed spectra, we created synthetic template spectra with the help of the stellar synthesis program {\tt SPECTRUM} \citep[][]{1994AJ....107..742G}, using ATLAS9 stellar models atmospheres \citep[][]{2004A&A...419..725C}. For each cluster we created the template spectrum from the theoretical atmosphere models that were closest in metallicity to the studied clusters and used the same spectral resolution as the observed spectra. Note that we did not derive or estimate the iron content from FLAMES data. From this data set we obtained the LOS velocity information for 77 cluster members for NGC~1851, and 26 members for NGC~4590. Our observation results are then combined with available FLAMES data\footnote{The proposals and the details of the existing FLAMES data can be found in \citet[][]{2018MNRAS.478.1520B}.}, resulting in data sets of 684 members for NGC~1851 and 263 members for NGC~4590. We compared the velocity of stars that in both our FLAMES/Giraffe and 2dF/AAOmega observations, finding that the two data sets have a systematic difference of $0.67\pm0.17\ \mathrm{km\ s^{-1}}$, with a intrinsic dispersion consistent with zero with the measurement uncertainties.

\subsection*{Metallicity}

Along with the spectral fitting of the data from 2dF/AAOmega, we integrated the spectrum template with the parameters from the last 1,000 steps of the MCMC chains to derive the equivalent width (EW), and adopt the median as the best-fitting value, and the 1$\sigma$ percentiles as uncertainties. The metallicity $\mathrm{[Fe/H]}$ of each star was estimated with the EW, the absolute magnitude of the Gaia G band, and the correlation from \citet{2020RNAAS...4...70S}. Since the metallicity estimation depends on the absolute magnitude, which is calculated based on the distance, this metallicity estimation is only valid for the member stars. The metallicity distribution of each cluster is fitted with a Gaussian distribution and the results are listed in Tab.~\ref{tab:data}. We note that the measured metallicity dispersion of the globular cluster could come from the under-estimation of the uncertainties of the metallicity of its member stars---we do not claim any intrinsic metallicity dispersion within the cluster members with our measurements, a topic which is beyond the scope of this work.

\begin{table*}
    \centering
    \begin{tabular}{|c|c|c|c|c|c|c|c|c|}
    \toprule
                & & &  \multicolumn{2}{c}{Number of} & \multicolumn{2}{c}{Measured metallicity} & \multicolumn{2}{c}{King truncation radius} \\
                            \cmidrule(lr){4-5} \cmidrule(lr){6-7} \cmidrule(lr){8-9} 
    Name        & D (kpc) & Peri-centre (kpc) &{\it good\_star} & members & Mean & Dispersion & Arcsec & pc   \\
    \hline
    NGC~1261 & 16.40 & $1.66\pm0.08$ & 71 & 44 & $-1.33\pm0.02$ & $0.11\pm0.02$ & 303.5 & 24.13 \\
    NGC~1851 & 11.95 & $0.87\pm0.04$ & 77 & 58 & $-1.19\pm0.02$ & $0.14\pm0.02$ & 391.2 & 22.66  \\
    NGC~1904 & 13.27 & $0.33\pm0.16$ & 97 & 48 & $-1.66\pm0.01$ & $0.10\pm0.01$ & 481.1 & 30.95 \\
    NGC~4590 & 10.40 & $8.89\pm0.07$ & 65 & 54 & $-2.32\pm0.02$ & $0.12\pm0.01$ & 894.5 & 43.93 \\
    \hline
    
    \end{tabular}
    \caption{The parameters of the target clusters. Column 1 is the name of the GCs. Column 2 is the heliocentric distance from \protect\citet{2021MNRAS.505.5957B}. Column 3 lists the peri-centre of their orbits within the Milky Way. Columns 4 and 5 show the number of {\it good\_star}s and selected members in each cluster. Columns 6 and 7 are the measured mean metallicity and dispersion. Columns 8 and 9 are the King truncation radius of each cluster from \protect\citet{Harris10}.}
    \label{tab:data}
\end{table*}

\section{N-body Simulations}
\label{sec:simulation}

To interpret the observations and estimate the influence of the external tidal field of the MW on the outer dynamical profiles of the GCs, we performed a series of direct $N$-body simulations in the following steps: we first integrated the orbit of the clusters backward in time for 2~Gyrs in the MW potential of \citet{Irrgangetal2013} using a fourth-order Runge-Kutta integrator, with the initial phase space parameters from \citet{baumgardtetal2019}, and the distance from \citet[][]{2021MNRAS.505.5957B}. We then set up $N$-body models of the clusters with different initial masses and sizes, which are non-rotating in an inertial reference frame. These initial $N$-body models were created based on the grid of $N$-body models described in \citet{2018MNRAS.478.1520B}. These models started from \citet{1962AJ.....67..471K} density profiles with different concentration parameters $c$, and followed a range of initial mass functions, starting with those from \citet{2001MNRAS.322..231K} and extending towards those that are more strongly depleted in low-mass stars. We then integrated the orbit of the clusters forward in time to the present-day position, using NBODY7 \citep{nitadoriaarseth2012} and NBODY6+P3T \citep{2022MNRAS.509.2075A} for the clusters NGC~1261 and NGC~4590 and NBODY6+P3T for NGC~1851 and NGC~1904, with the MW potential of \citet{Irrgangetal2013} for an external tidal field in both codes\footnote{Simulations are run on the OzSTAR GPU cluster of Swinburne University and the GPU cluster of the University of Queensland.}. Note that the initial cluster models of \citet{2018MNRAS.478.1520B} were unsegregated, however, mass segregation developed dynamically over time, so the simulations presented here started from already mass segregated models. 

\begin{figure}
    \centering
    \includegraphics[width=\columnwidth]{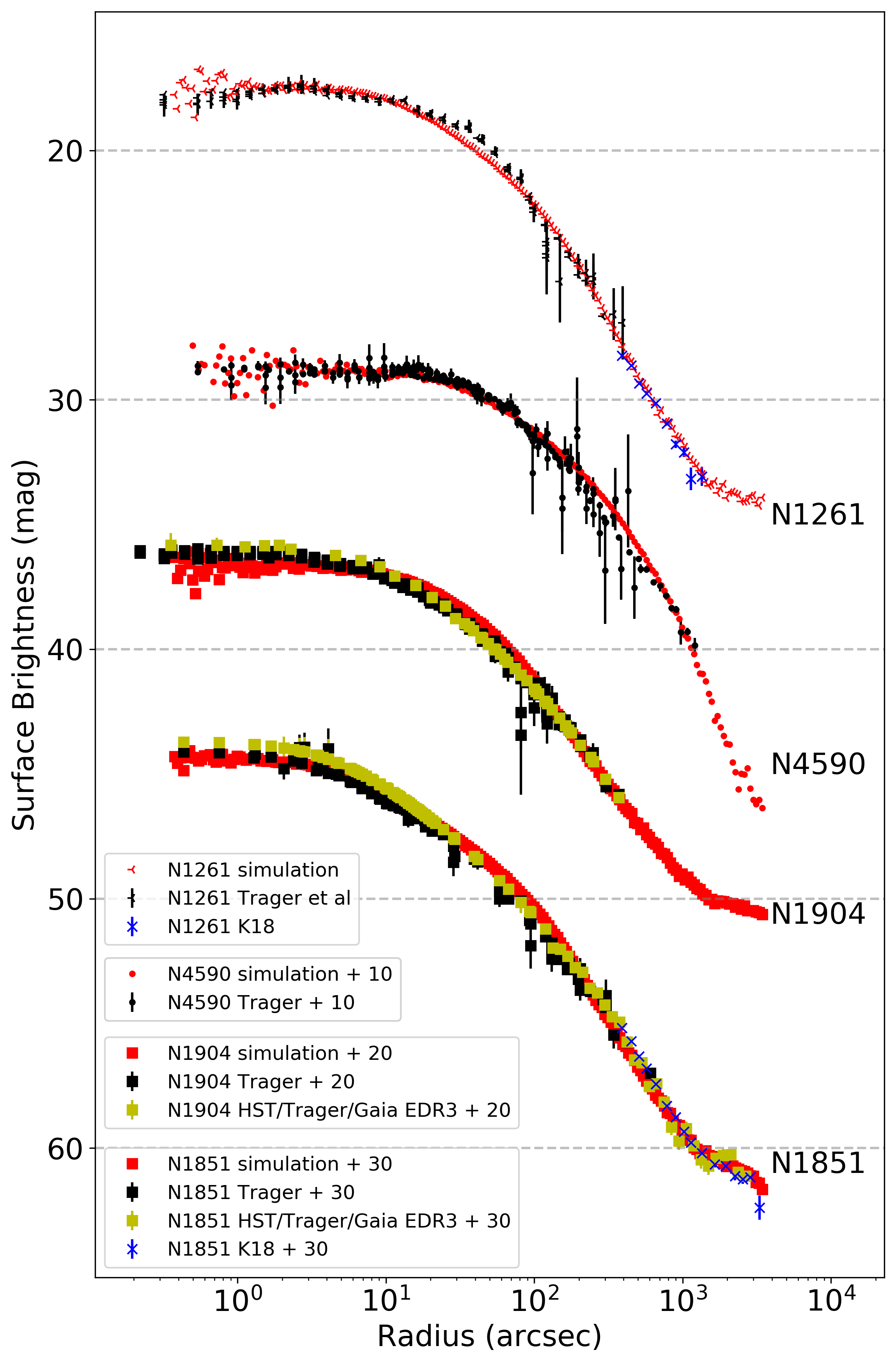}
    \caption{The comparison of the observed surface brightness profiles with the simulations after 2 Gyr of evolution. The profiles of NGC~4590, NGC~1904 and NGC~1851 are shifted to have a better presentation. The observational data come from \protect\citet{1995AJ....109..218T}. For NGC~1904 and NGC~1851, we also add a combination of data from HST, Gaia EDR3 from \protect\citet[][]{2020PASA...37...46B} as a supplement to the inner profile from \protect\citet{1995AJ....109..218T}. We also include the profiles from \protect\citet[][K18]{2018MNRAS.473.2881K} for NGC~1261 and NGC~1851.
    }
    \label{fig:SBP}
\end{figure}

\begin{figure}
    \centering
    \includegraphics[width=\columnwidth]{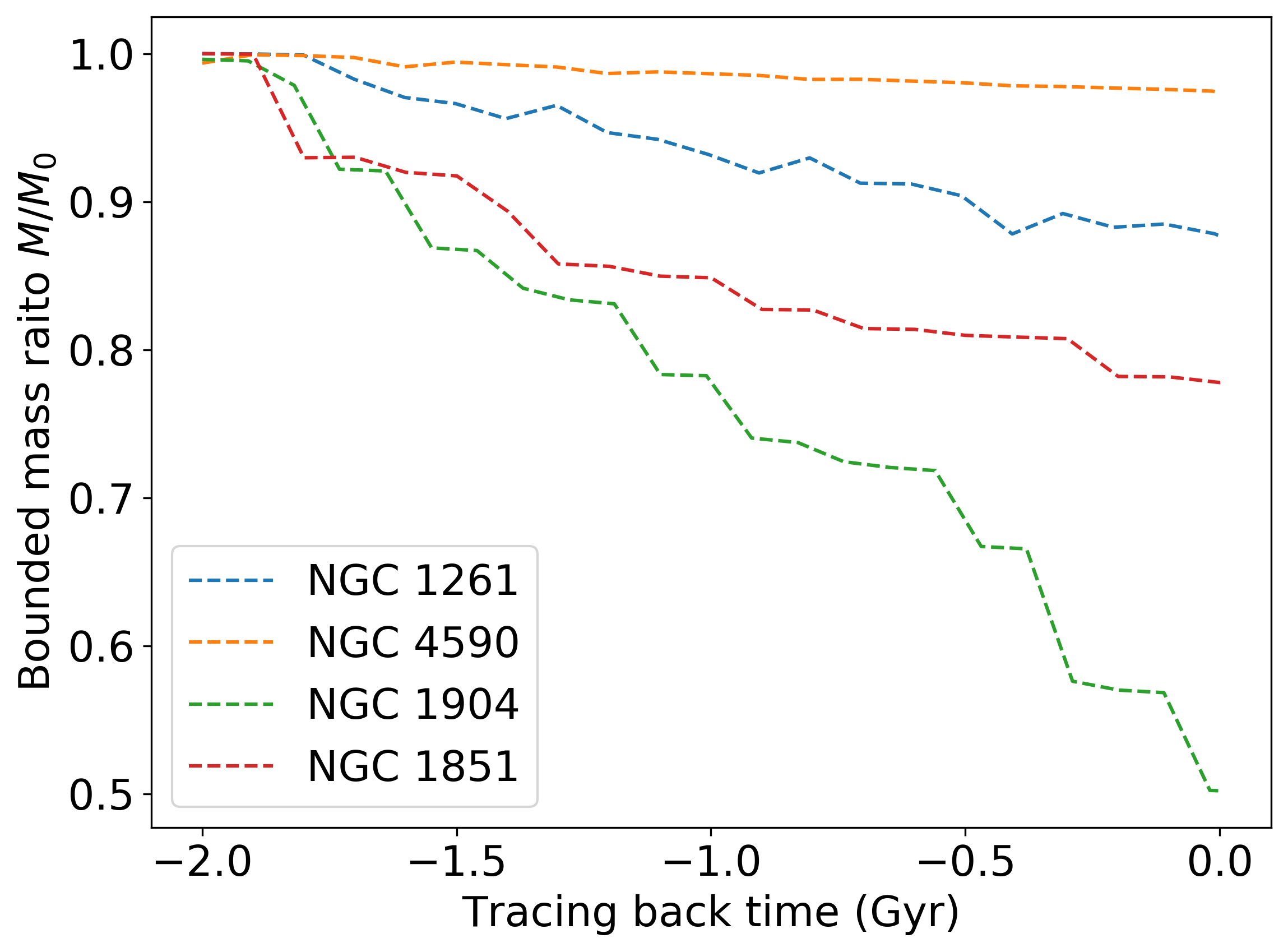}
    \caption{The bound mass evolution of the clusters in the simulation within 2 Gyr, where the mass is normalised to the starting mass $\mathrm{M_0}$. Compared to NGC~1261 and NGC~4590, NGC~1851 and NGC~1904 have a much larger mass loss rate due to stronger interaction with the MW.
    }
    \label{fig:bound_mass}
\end{figure}

We select the best-fitting model based on the comparison to the surface brightness profiles, the observed velocity dispersion profile and the mass function data at different radii. Fig.~\ref{fig:SBP} shows the surface brightness profile of the simulation and the observational data, where the main observational dataset is from \citet{1995AJ....109..218T}, and as supplements to the inner profiles, we add the profiles from \citet[][K18]{2018MNRAS.473.2881K} for NGC~1261 and NGC~1851, and the data from HST and Gaia EDR3 for NGC~1904 and NGC~1851 from \citet[][]{2020PASA...37...46B}. The simulations for NGC~1261 and NGC~4590 match the observed surface brightness profiles well, with only a deviation around $\sim 300\ \mathrm{arcsec}$ in the NGC~1261 profile. Though NGC~1904 and NGC~1851 have intense interaction with the MW due to their close pericentres, which makes it difficult to simulate the clusters, the simulations still show reasonable agreement with the observations.

The simulations are also compared to the observed streams associated with the clusters NGC~1261, NGC~4590 and NGC~1851, as shown in the appendix and Fig.~\ref{fig:sim_stream}. Though NGC~1851 shows a discrepancy, where the track deviates from the ridge of the stellar distribution in the simulation, we find that our simulations match well with the streams associated with NGC~1261 and NGC~4590---for NGC~4590, the simulation reproduces the stream track over a large spatial extent. 

Fig.~\ref{fig:bound_mass} shows the bound mass evolution of the clusters in the simulations. NGC~1261 lost $12.3\%$ of its starting mass within the 2 Gyr simulation, and NGC~4590 lost $2.7\%$ of its initial mass. NGC~1904 has a much larger mass loss rate---$\sim50\%$ within 2 Gyr---since it has a very small pericentre ($\le 1\ \mathrm{kpc}$) similar to NGC~1851, which loses $\sim 29\%$ within 2 Gyr. Given the bound mass evolution, and assuming a linear mass loss rate \citep[see e.g.,][]{2003MNRAS.340..227B}, NGC~1261 would have lost $\sim84\%$ of its current mass in $12$ Gyr of evolution due to the tidal interaction, and will be fully destroyed in $\sim14.2$ Gyr; NGC~4590 only lost $\sim 17\%$ of its current mass in $12$ Gyr of evolution; NGC~1904 would have lost more than $6.0$ times its current mass due to the strong tidal interaction; and NGC~1851 lost $1.7$ times its current mass within $12$ Gyr due to tidal interaction. The $N$-body simulation does not include stellar evolution, but note that the mass loss from stellar evolution mainly occurs in the early evolution of the cluster, and finishes before other mechanisms become important, losing about half of the initial total mass \citep[e.g.,][]{2003MNRAS.340..227B,2010MNRAS.409..305L}.

\section{Results}
\label{sec:results}

\begin{table}
    \centering
    \begin{tabular}{c|c|c|c|}
        \hline
        Cluster & Rot. ($\mathrm{km\ s^{-1}}$) & $v_{sys}$ ($\mathrm{km\ s^{-1}}$)& $\sigma_{v_{los}}$ ($\mathrm{km\ s^{-1}}$)\\
        \hline
        NGC~1261 & $1.66^{+0.88}_{-0.92}$ & $73.04^{+0.52}_{-0.50}$ & $3.22^{+0.45}_{-0.38}$ \\
        NGC~1851 & $0.34^{+1.28}_{-1.38}$ & $322.17^{+0.69}_{-0.71}$ & $4.64^{+0.57}_{-0.47}$\\
        NGC~1904 & $-0.88^{+3.35}_{-3.90}$ & $206.48^{+2.08}_{-2.02}$ & $7.52^{+2.18}_{-1.51}$ \\
        NGC~4590 & $2.05^{+6.91}_{-3.67}$ & $-92.07^{+1.60}_{-1.42}$ & $2.14^{+2.61}_{-1.15}$\\
        \hline
    \end{tabular}
    \caption{The estimated integrated dynamical properties of each cluster from the AAT 2df/AAOmega data. The first column is the name of the clusters; the second column shows the rotation amplitude; the third column shows the systemic LOS velocity, and the last column is the LOS velocity dispersion.
    }
    \label{tab:observation_properties}
\end{table}

In this section, we will discuss the dynamical results of NGC~1261, NGC~4590, NGC~1904, and NGC~1851, based on the LOS velocities of the members. With the velocity and metallicity from the reduced data, we can select members from the targets and exclude foreground/background stars. For each star, the perspective view effect is calculated based on the projected position and the systemic velocity---the LOS velocity, proper motion, and the distance from \citet{2018MNRAS.478.1520B}---with Eq.~6 of \citet{2006A&A...445..513V}. Then we apply the following velocity clipping for the members, where the $\Delta$ below indicate the relative velocities with respect to the clusters' systemic values: 

\begin{gather}
    -25 < \Delta PM_{R.A.} < 25\ \mathrm{km\ s^{-1}} \notag \\
    -25 < \Delta PM_{Dec.} < 25\ \mathrm{km\ s^{-1}} \notag \\
    -25 < \Delta v_{los} < 25\ \mathrm{km\ s^{-1}} 
    \label{eq_selection}
\end{gather}

We note that with the distance to each GC, the proper motions are converted into velocities, and the uncertainties are smaller than $10\ \mathrm{km\ s^{-1}}$ for most ($> 90\%$) of the {\it good\_star}s. However, compared to the spectral redshift, the uncertainties from proper motion are significantly larger and the number of adopted members for each cluster is small, hence the proper motions are only used to eliminate the contaminants without further analysis. The $\pm 25\ \mathrm{km\ s^{-1}}$ selection limit is significantly larger than the velocity dispersion and rotation amplitude of the clusters (see the following sections), and so minimises the influence of the velocity clipping on the final results. Combined with the constraints from the CMD when we select the targets, this velocity clipping leads to a clean sample of members. We examined the background level of each cluster by counting the stars with $-50 < \Delta v_{x} < 50\ \mathrm{km\ s^{-1}}$ (here the $x$ represent the velocity in $R.A.$, $Dec.$ and LOS direction) that are not identified as members by Eq.~\ref{eq_selection}. We assume a smooth distribution of the background stars in the velocity space and find the expected number of background stars within the velocity range from Eq.~\ref{eq_selection}. The results for NGC~1261 is $0.71$; for NGC~1904 is $0.29$; for NGC~1851 is $0.14$ and for NGC~4590 is $0$. Hence we expect a low background level in our samples. The same velocity selections will also be applied to the simulations. In Sec.~\ref{sec:results}, we will present the profiles of the simulations with and without the velocity clipping.

\begin{figure}
    \centering
    \includegraphics[width=\columnwidth]{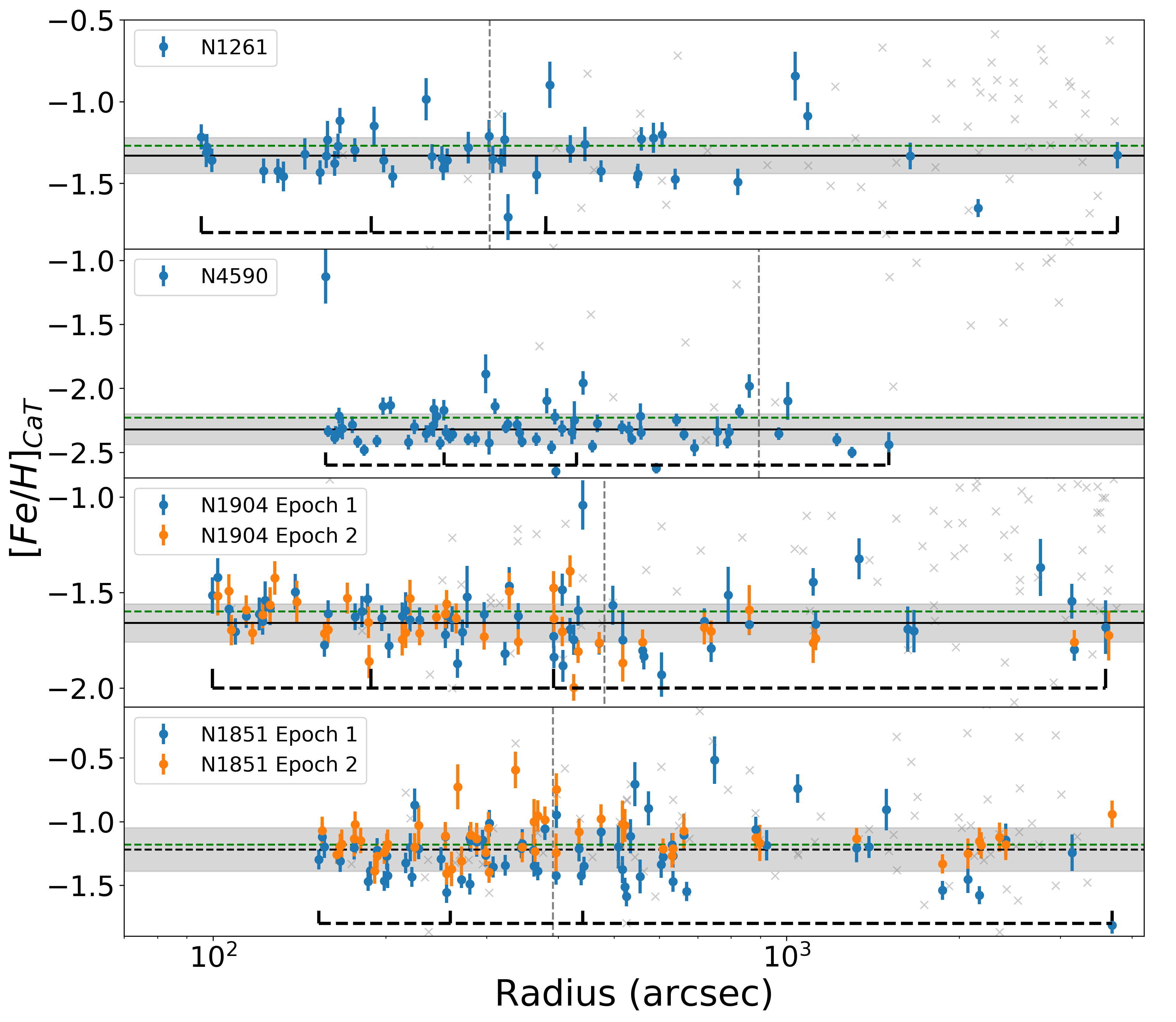}
    \caption{The metallicity of the four clusters derived from the CaT EW distributed along the radius. The points with error bars are cluster members that follow the velocity clipping, while the grey points are stars that were excluded. The green dashed lines show the metallicity from \protect\citet{Harris10}, the black lines indicate the best-fitting mean metallicity and the grey regions show the dispersion from the fitting. The vertical grey dashed lines show the truncation radius of each cluster and the black dashed lines indicate the radial range of the bins used in the dynamical analyses. 
    }
    \label{fig:Feh}
\end{figure}

\begin{figure}
    \centering
    \includegraphics[width=\columnwidth]{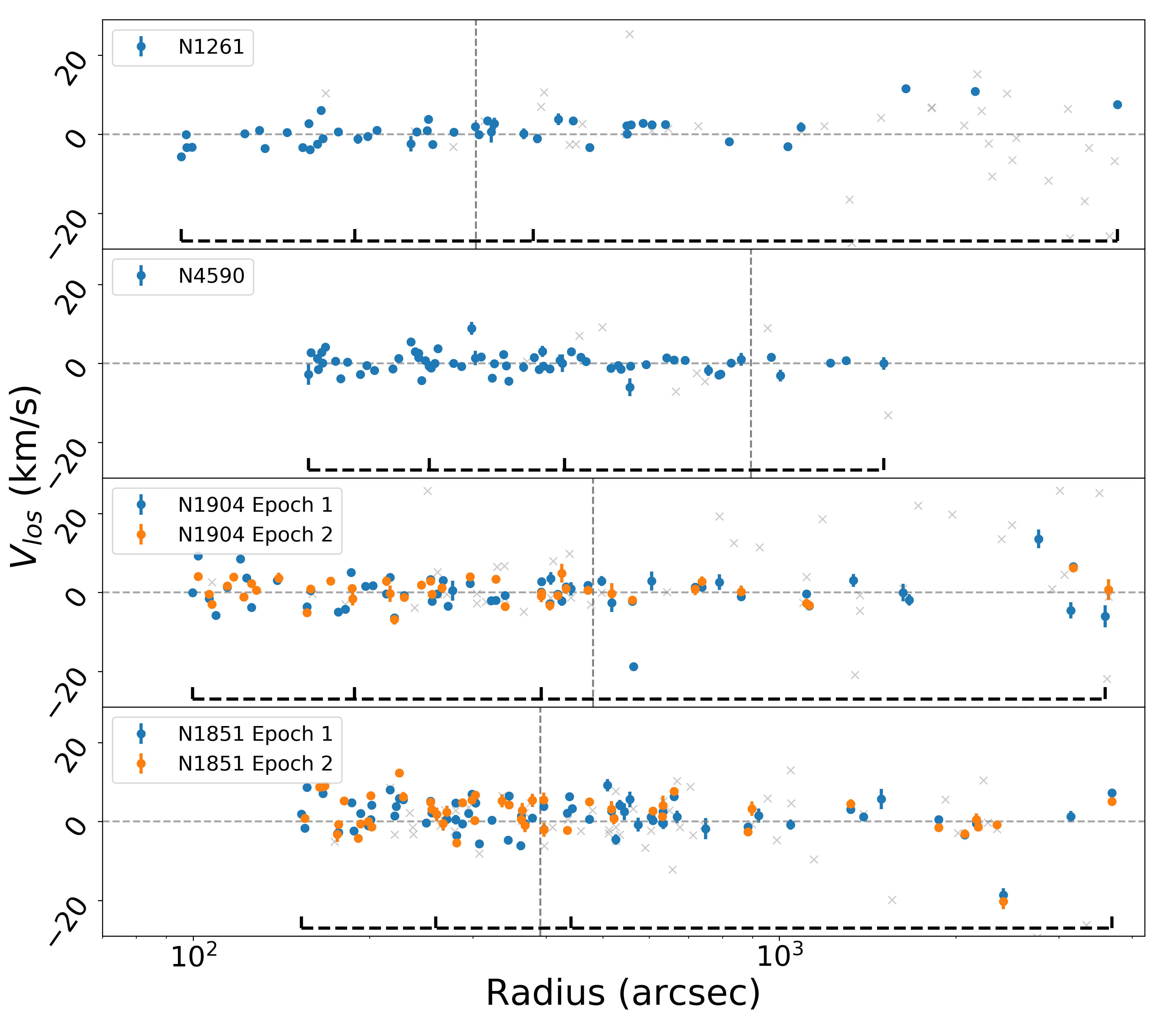}
    \caption{The LOS velocity distribution of the cluster members (the perspective view effect and the systemic velocity have been subtracted), where the grey crosses are excluded targets. The black dashed lines at the bottom of each panel show the boundary of the radial bins that are used in the dynamical analyses.The vertical grey dashed lines show the truncation radius of each cluster and the horizontal grey dashed lines indicate $0\ \mathrm{km\ s^{-1}}$ for reference.}
    \label{fig:VLOS}
\end{figure}

Fig.~\ref{fig:Feh} shows the metallicity of the selected members compared to the metallicity from \citet{Harris10} (the green dashed lines), while the mean and one sigma range of the metallicity distributions are shown by black dashed lines and grey regions. Tab.~\ref{tab:data} also lists the measured mean and dispersion of the metallicity distribution. The results show that the members selected from the velocity clipping have a narrow metallicity distribution, suggesting that the selection from Eq.~\ref{eq_selection} is effective. Comparing to \citet{Harris10}, we found that the measurements of the metallicity agree with the literature (within an $1\sigma$ uncertainty). Fig.~\ref{fig:VLOS} shows the velocity distribution of the selected members, where the stars are concentrated around $v_{\mathrm{los}} = 0\ \mathrm{km\ s^{-1}}$ for all clusters, suggesting the samples are clean. The spatial distribution of the selected stars from the observations and the simulations, as well as the dynamical analysis of the members, are presented in the corresponding subsections.

To fit for rotation and velocity dispersion with the selected sample we fit a simple model, 
\begin{gather}
    v_{\mathrm{los}, 0} = A_{\mathrm{rot}} \sin (\phi - \phi_0) + v_{\mathrm{sys}} \notag \\
    p(v_{\mathrm{los}}) = \frac{1}{\sqrt{2\pi}\sigma_{v_{\mathrm{los}}}} \exp^{-\frac{(v_{\mathrm{los}} - v_{\mathrm{los}, 0})^2}{2\sigma_{v_{\mathrm{los}}}^2}},
    \label{eq:model1}
\end{gather}
to the data in different radial bins with the measured velocity uncertainties and the $0.99\ \mathrm{km\ s^{-1}}$ systematic uncertainty applied. Here the $v_{\mathrm{los}, 0}$ is the expected LOS velocity from this model, $A_{\mathrm{rot}}$ represents the rotational amplitude, $\phi$ is the direction angle on tangent plane around the cluster and $\phi_0$ is the reference angle along which direction the rotation axis follows (here the position angle is defined as from North ($0^{\circ}$) to East ($90^{\circ}$)). The $v_{\mathrm{sys}}$ is systemic velocity, which is converted to the corresponding value based on the position of each star. We assume that the observed LOS velocity follows a Gaussian distribution centred on the expected velocity $v_{\mathrm{los}, 0}$, where the $p(v_{\mathrm{los}})$ is the probability of observing $v_{\mathrm{los}}$ given that $\sigma_{v_{los}}$ is the intrinsic dispersion of the GC. Tab.~\ref{tab:observation_properties} lists the integrated properties of each globular cluster estimated by fitting the whole data set from 2dF/AAOmega with Eq.~\ref{eq:model1}. In the following subsections, we group the members of each cluster into the radial bins shown as the black dashed lines in Fig.~\ref{fig:Feh}, Fig.~\ref{fig:VLOS} and in the following figures for each cluster, where each radial bin in each cluster have the same number of members, and discuss the dynamical features including the dispersion and rotation of each GC. We use bins in order to estimate the dynamical properties of the clusters, and the numbers of members in each bin are same in each cluster. We note that the binned data can lead to artefacts. For example, the first data bin in NGC~1261 shows unexpected low dispersion and large rotation. This is more likely to be a binning effect.

We also include different data sets as comparisons to our observations---the data from the $N$-body simulations and the observational data from existing literature in the inner part are shown in the following figures (the references for the LOS velocity dispersion profiles are cited in the appropriate cluster subsections). In addition, we include the theoretical predictions from {\sc limepy} \citep{2015MNRAS.454..576G} and {\sc spes} \citep{2019MNRAS.487..147C} models from \citet{2019MNRAS.485.4906D}, where the GC masses are adopted from \citet{2018MNRAS.478.1520B}. These models elaborate on the King and Wilson models to include radial velocity anisotropy, where the {\sc spes} model allows for a description of stars near the escape energy including the effect of marginally unbound stars (the PEs). The detailed parameters used in the {\sc limepy} and {\sc spes} models are presented in Tab.~\ref{tab:model_parameter}.

\begin{table*}
    \centering
    \begin{tabular}{|c|c|c|c|c|c|c|c|c|}
    \toprule
                &  &  \multicolumn{3}{c}{{\sc limepy} Model} & \multicolumn{4}{c}{{\sc spes} model} \\
                            \cmidrule(lr){3-5} \cmidrule(lr){6-9} 
    Name        &  Mass($10^5\mathrm{M_{\odot}}$) & $W_0$ & g & $r_h$ (pc) & $W_0$ & $\eta$ & $\mathrm{log_{10}(1-B)}$ & $r_h$ (pc)  \\
    \hline
    N1261     & $1.67\pm0.17$ & $3.63\pm0.41$  & $2.82\pm0.12$ & $4.33\pm0.07$ & $4.99\pm0.10$ & $0.23\pm0.01$ & $-2.59\pm0.22$ & $4.45\pm0.04$  \\
    N4590     & $1.23\pm0.12$ & $5.17\pm0.08$ & $2.46\pm0.04$ & $5.74\pm0.05$ & $5.74\pm0.06$ & $0.22\pm0.01$ & $-2.63\pm0.17$ & $5.86\pm0.06$  \\
    N1904     & $1.69\pm0.11$ & $6.79\pm0.20$ & $1.89\pm0.09$ & $3.20\pm0.10$ & $6.57\pm0.07$ & $0.13\pm0.08$ & $-3.94\pm1.69$ & $3.17\pm0.12$  \\
    N1851     & $3.02\pm0.04$ & $7.64\pm0.06$ & $1.85\pm0.02$ & $2.51\pm0.05$ & $7.46\pm0.05$ & $0.13\pm0.01$ & $-2.64\pm0.13$ & $2.43\pm0.06$  \\ 
    \hline
    
    \end{tabular}
    \caption{The parameters used in the {\sc limepy}/{\sc spes} models. The GC mass is taken from \citet{2018MNRAS.478.1520B}, and the model parameters are adopted from \citet{2019MNRAS.485.4906D}. $W_0$ indicates the concentration of the cluster (the dimensionless central potential); $g$ in {\sc limepy} model describes the sharpness of the truncation in energy; $\eta$ in {\sc spes} models represents the ratio of the velocity dispersion of the PEs over a velocity scale, which sets the physical scales of the model; the parameter $B$ in {\sc spes} model constrains the amount of PEs, where $B=1$ represents no PEs; $r_h$ in each model is the half-mass radius.  }
    \label{tab:model_parameter}
\end{table*}

\subsection*{NGC~1261}

\begin{figure*}
    \centering
    \includegraphics[width=2\columnwidth]{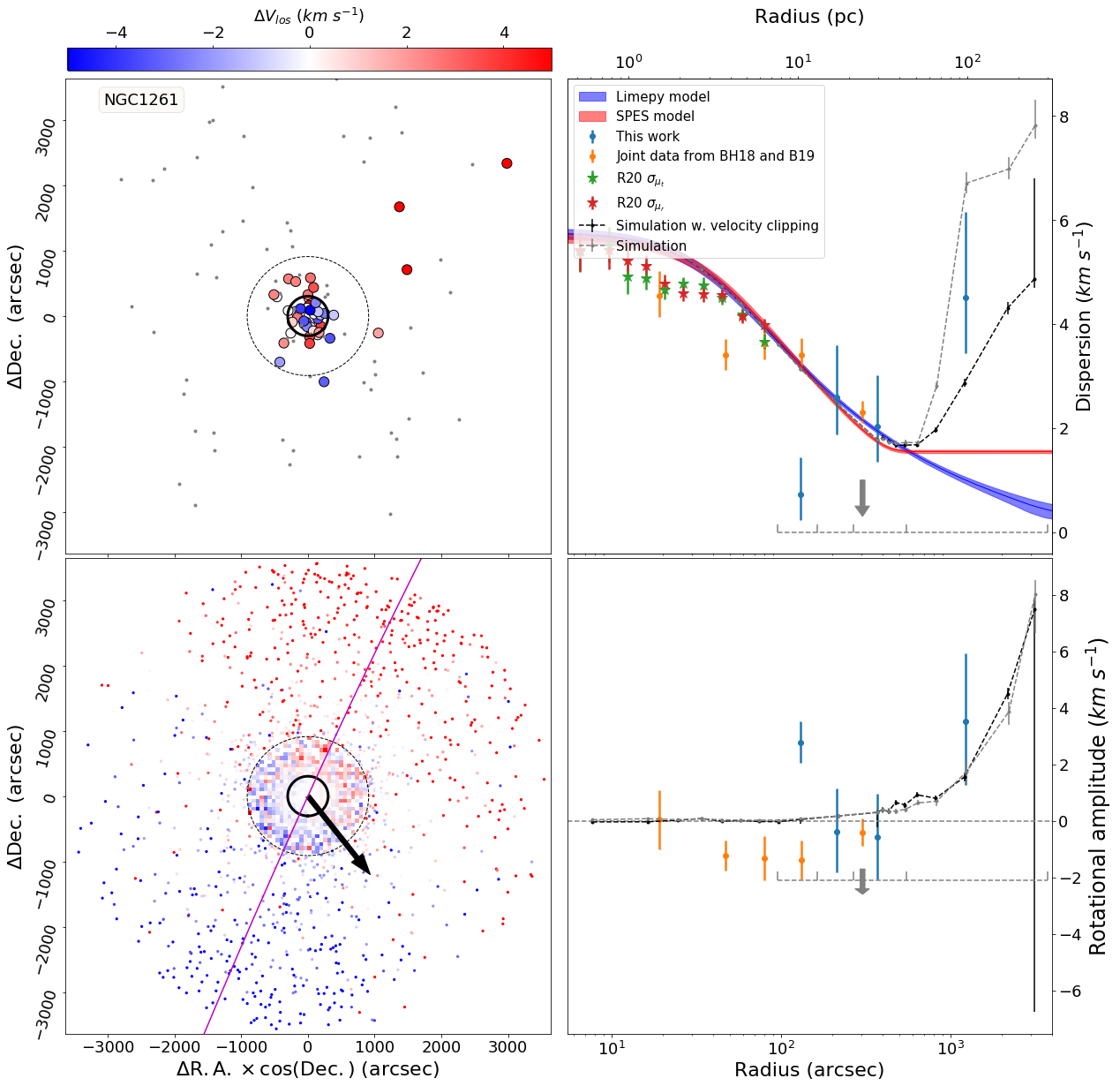}
    \caption{{\it Left:} The distribution of NGC~1261 stars from the observations (top) and the simulation (bottom), where the colour indicates the LOS velocity with the perspective view effect subtracted. The solid black circles show the King tidal radius. The dashed black circle in the simulation panel indicates the dense region where it is difficult to distinguish single stars. Hence we use the heat map to replace the scatter points within the dashed circle for the simulation, where the colour indicates the average LOS velocity within each bin. For NGC~1261, the radius of the dashed circle is $3r_{\rm t}$ (here $r_{\rm t}$ is the truncation radius from the King model fit and is the same in the following figures). The magenta line in the bottom left simulation panel indicates the stream track from \protect\citet{2022arXiv220410326M} that is shifted to match the cluster centre, where the stream data are from \protect\citet[][]{2021ApJ...914..123I} (see also the figure in the appendix). The black arrow in the bottom left panel indicates the direction of the systemic velocity of the cluster, and the grey points in the top left observation panel are the excluded stars.\\
    {\it Right:} The results of the dynamical fitting for NGC~1261. The blue points with error bars are the data from our observations. Here we also include some published data covering the inner part of the cluster---the joint data from \protect\citet[][BH18]{2018MNRAS.478.1520B} and \protect\citet[][B19]{baumgardtetal2019} (the orange error bars), and \protect\citet[][R20]{2020ApJ...895...15R} (the green and the red error bars). The blue and red lines represent the {\sc limepy} model and the {\sc spes} model respectively \protect\citep{2019MNRAS.485.4906D}, while the blue and red regions are the $1\sigma$ uncertainties of the models. The dash-dotted lines are the profiles from the $N$-body simulation (the black line is the profile with the velocity clipping Eq.~\ref{eq_selection}, and the gray line is the profile without any selection). The horizontal grey dashed lines indicate the radial bins of our observations. {\it Top:} The dispersion profiles from the literature and from our observations, where the grey arrow indicates the truncation radius of the best-fitting King model from \protect\citet{Harris10}, beyond which the dispersion profiles of both the simulation and observations increase significantly in the outer parts. {\it Bottom:} The rotation profile of NGC~1261 from the observations and the simulation. The grey dashed line represents the zero rotational amplitude as a reference.
    }
    \label{fig:N1261_results}
\end{figure*}

Our first target is NGC~1261, which has been found to have a stellar envelope that extends out to $\sim 22\,\, \mathrm{arcmin}$ \citep[e.g.][]{2012MNRAS.419...14C,2018MNRAS.473.2881K}. A long tidal structure that is associated with NGC~1261 was identified by \citet{2021ApJ...914..123I}. The present-day mass of NGC~1261 is $1.67\times10^5\ \mathrm{M_{\odot}}$ \citep[][]{2018MNRAS.478.1520B}, and it is currently located close to the apocentre of its orbit around the Galactic centre \citep{2014MNRAS.442.1569W,2017MNRAS.464.2174B}. Based on the simulation, the orbital period of NGC~1261 is about $\sim 170\ \mathrm{Myr}$, with a pericentre radius of $\lesssim 2\ \mathrm{kpc}$, and an apocentre radius of $\sim 20\ \mathrm{kpc}$. Given this short orbital period, the cluster completed more than 10 orbits around the Galactic centre within the $2\ \mathrm{Gyr}$ of our simulations. The frequent interaction with the  tidal field of the inner Galaxy generates strong tidal tails around the GC in the simulation, which extend predominantly in a North to South direction (see the bottom left panel in Fig.~\ref{fig:N1261_results}).

In this survey, NGC~1261 has a single observation, with 138 spectra taken for stars within 1 degree of the GC centre, from which we have 73 stars classified as {\it good\_star}, with 46 stars selected with the velocity clipping from Eq.~\ref{eq_selection}, and we exclude 2 stars brighter than 14 mag (see Sec.~\ref{sec:data}). In the top panel of Fig.~\ref{fig:Feh} we present the metallicity of the members selected based on the dynamics. We estimate the mean metallicity of our sample is $\mathrm{[Fe/H]} = -1.33 \pm 0.02$ with a dispersion of $0.11 \pm 0.02$. The narrow distribution of the metallicity suggests that the sample is clean, while the faint stars have larger metallicity uncertainties and consequently spread over a larger range.

The top left panel of Fig.~\ref{fig:N1261_results} shows the distribution of the selected members, where the colour indicates the LOS velocity with the perspective view effect subtracted \footnote{In the following discussion, unless specified otherwise, all velocity data represents the velocity with the perspective view effect subtracted, i.e. velocities are in the co-moving frame with the GC.} The bottom left panel shows the corresponding simulation, and the colour indicates the LOS velocity as well. The inner part of the simulation (indicated by the dash circle) is too crowded to effectively show the LOS velocity pattern, hence we replace the scatter points with a heat map to present the average LOS velocity within $3r_{\rm t}$ ($r_{\rm t}$ is the truncation radius of the best-fitting King model from \citet[][]{Harris10}) with the same colour scheme. Within the truncation radius, the simulated cluster is dominated by random motion---the average LOS velocity is close to zero. The outer part appears to be rotating due to the effect of the tidal tails, where the south-western part has negative velocity and the north-eastern part has positive velocity. The stars from both the observations and the simulation extend well beyond the truncation radius of King model \citep{Harris10}, with the most distant member in the observations being $\sim 0.8$ degree away from the GC centre. 

We estimated the observed dispersion and rotation with Eq.~\ref{eq:model1}, and the results are shown in the right panels of Fig.~\ref{fig:N1261_results}. The dispersion derived from our observations appears to decrease within the truncation radius and increases beyond that. The innermost bin shows significantly lower dispersion, which might be a false detection due to the lack of data points, where part of the velocity dispersion is explained as regular rotation by the kinematic fit. In the figure, we also include the dispersion profiles from \citet[][R20]{2020ApJ...895...15R}, and the joint data from \citet[][here after BH18]{2018MNRAS.478.1520B} and \citet[][here after B19]{baumgardtetal2019}, in the inner part extending to $\sim 5$ arcsec from the centre. The observational data sets are in agreement with each other in the overlap regions (except for the innermost bin in our observations). As comparisons, we also include the predictions from the {\sc limepy} and {\sc spes} models and the dispersion and the rotation profile derived from the simulations. The simulation and the models basically predict the same dispersion profile within the truncation radius, while the simulation suggests a much larger and increasing dispersion profile beyond that. The velocity clipping Eq.~\ref{eq_selection} excludes stars with large relative velocity, and leads to a lower dispersion profile. Within the truncation radius, the simulation and the {\sc limepy} and {\sc spes} models agree well with our observation (except for the discrepancies around $10-40$ arcsec). In the last radial bin, the observed dispersion is $4.51^{+1.64}_{-1.07}\ \mathrm{km\ s^{-1}}$, and the nearest bin in the simulation yield a dispersion of $2.87\pm0.08\ \mathrm{km\ s^{-1}}$ lower than the observation.

In the simulations, the cluster has no rotation signal within the truncation radius, but the escaped stars around the cluster lead to a clear increasing rotation signal beyond that. In our observations, the mean rotation in the outermost bin is $3.52^{+2.24}_{-2.41}\ \mathrm{km\ s^{-1}}$ with a position angle of $87^{\circ}.1^{+48.1}_{-48.7}$. Though larger than the simulation prediction, the difference is within the uncertainty (the closest simulation radial bin has an amplitude of $1.73^{+0.37}_{-0.41}\ \mathrm{km\ s^{-1}}$ with a position angle of $111^{\circ}.0^{+19.0}_{-12.6}$). The large amplitudes (of both dispersion and rotation) in the last bin mainly come from three stars with the LOS velocity of $(10.87\pm0.57, 7.51\pm1.10, 11.54\pm0.55)\ \mathrm{km\ s^{-1}}$, and the corresponding metallicity of $(-1.65\pm0.06, -1.32\pm0.08, -1.33\pm0.08)\ dex$. If we exclude the first star that has a metallicity of $-1.65$ (which suggests that this star is unlikely to be a member), the resultant dispersion is $4.14^{+1.60}_{-1.06}\ \mathrm{km\ s^{-1}}$, and the rotation amplitude is $3.29^{+2.12}_{-2.25}\ \mathrm{km\ s{^-1}}$. We see some non-zero rotation signal in the inner part of the cluster, however, as we noted above, this signal is likely due to the lack of data points. If we include the two stars brighter than $14$ mag (see Sec.~\ref{sec:data}), the rotation signal is consistent with zero.

\subsection*{NGC~4590}

\begin{figure*}
    \centering
    \includegraphics[width=2\columnwidth]{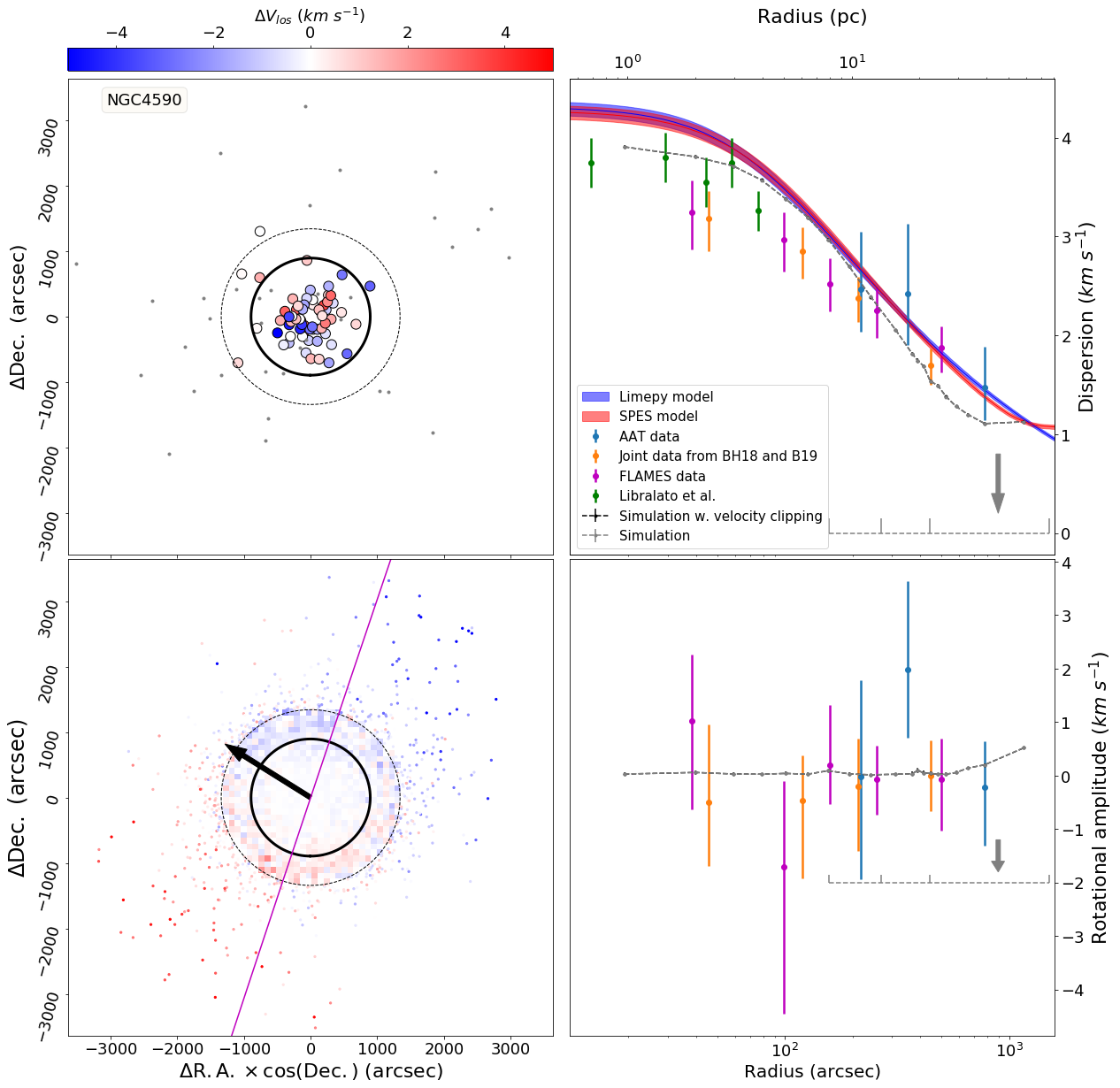}
    \caption{The same figure as Fig.~\ref{fig:N1261_results} for NGC~4590. Here the stream data is from \protect\citet{2019MNRAS.488.1535P}, and the radius for the dashed circle is $2r_{\rm t}$ within which we replace the scatters with the heat map. In the right panel, we include the joint data from BH18/B19 (orange error bars) and the data from FLAMES observation (magenta error bars), and the data from \protect\citet[][]{2022ApJ...934..150L} (green error bars) to show the profiles in the inner parts.
    }
    \label{fig:N4590_results}
\end{figure*}

NGC~4590 is among the most metal-poor and oldest GCs in the MW \citep[][]{Harris10}. It has a present-day mass of $1.23\times10^{5}\ \mathrm{M_{\odot}}$ \citep[][]{2018MNRAS.478.1520B}, and a heliocentric distance of $\sim 10.4\ \mathrm{kpc}$, and a galactocentric distance of $\sim 10.3\ \mathrm{kpc}$. The simulation shows that NGC~4590 is close to its pericentre, with an orbital period of $\sim 400\ \mathrm{Myr}$ and the pericentre of $\sim 9\ \mathrm{kpc}$. The tidal interaction between the cluster and the MW appears to leave a signal on this cluster: while a tidal tail is barely detected close to the cluster \citep[e.g.][]{2014MNRAS.445.2971C,2020MNRAS.495.2222S}, NGC~4590 is associated with the Fj\"orm stream based on their orbital properties \citep{2019MNRAS.488.1535P,2021ApJ...914..123I,2021ApJ...909L..26B}.

From the observations, we extracted 100 {\it good\_star} spectra, from which 54 targets were selected as members and used in the dynamical analyses. The second panel of Fig.~\ref{fig:Feh} shows the metallicity of the selected members. Compared to \citet{Harris10}, we estimate a lower metallicity of $\mathrm{[Fe/H]} = -2.34 \pm 0.02$, with a dispersion of $0.10 \pm 0.01$. One star at the centre of the cluster shows a significantly higher metallicity, which is excluded from the following analysis. The left panels of Fig.~\ref{fig:N4590_results} present the distribution for the observations and the simulation, respectively, where the colour indicates the LOS velocity. Cluster members in both the simulation and observations are found beyond the truncation radius of the King model (the black circle), but compared to NGC~1261, the cluster stars are more concentrated towards the GC centre. In the outskirts beyond the truncation radius, the LOS velocity of the simulation shows a pattern that the north-eastern side has a smaller velocity than the south-western side, leading to a rotation signal in the outskirts (see the bottom right panel of Fig.~\ref{fig:N4590_results}).

We calculated the dispersion and the rotation using Eq.~\ref{eq:model1}, and the results are shown in the right panels of Fig.~\ref{fig:N4590_results}. Though we do find members beyond the truncation radius, the average radii in all the bins are smaller than the truncation radius. The dispersion profile from our observation shows a decreasing tendency outwards, and agrees well with the results from BH18/B19 , as well as the FLAMES data in the overlap regions. The dispersion profile derived from our observations follows the predictions from the {\sc limepy} model and the {\sc spes} model, though the BH18/B19 data suggests a smaller dispersion, especially in the inner part ($< 200$ arcsec) of the cluster. The simulation shows a different dispersion profile, which agrees well with the profile from \citet[][]{2022ApJ...934..150L}, but over-estimates the dispersion from BH18/B19 and FLAMES within $\sim 50$ arcsec, and under-estimates the dispersion within the last radial bin. The velocity selection does not effectively change the dispersion and the rotation profiles, which mostly overlap with the profiles of the simulation without velocity clipping. Our results, as well as the BH18/B19, are only marginally inconsistent with the simulation beyond $200$ arcsec. Most of the rotational measurements (from 2dF/AAOmega, FLAMES and BH18/B19) are consistent with zero, in agreement with previous studies reporting non-detections of rotation in this cluster \citep[e.g.,][]{2010MNRAS.406.2732L,2015AJ....149...53K,2018MNRAS.481.2125B,2019MNRAS.484.2832V}. In the middle bin of 2dF/AAOmega measurement, we see a rotational signal of $1.99^{+1.65}_{-1.28}\ \mathrm{km\ s^{-1}}$ with a position angle of $39^{\circ}.6^{+22.9}_{-62.5}$.

\subsection*{NGC~1904}

\begin{figure*}
    \centering
    \includegraphics[width=2\columnwidth]{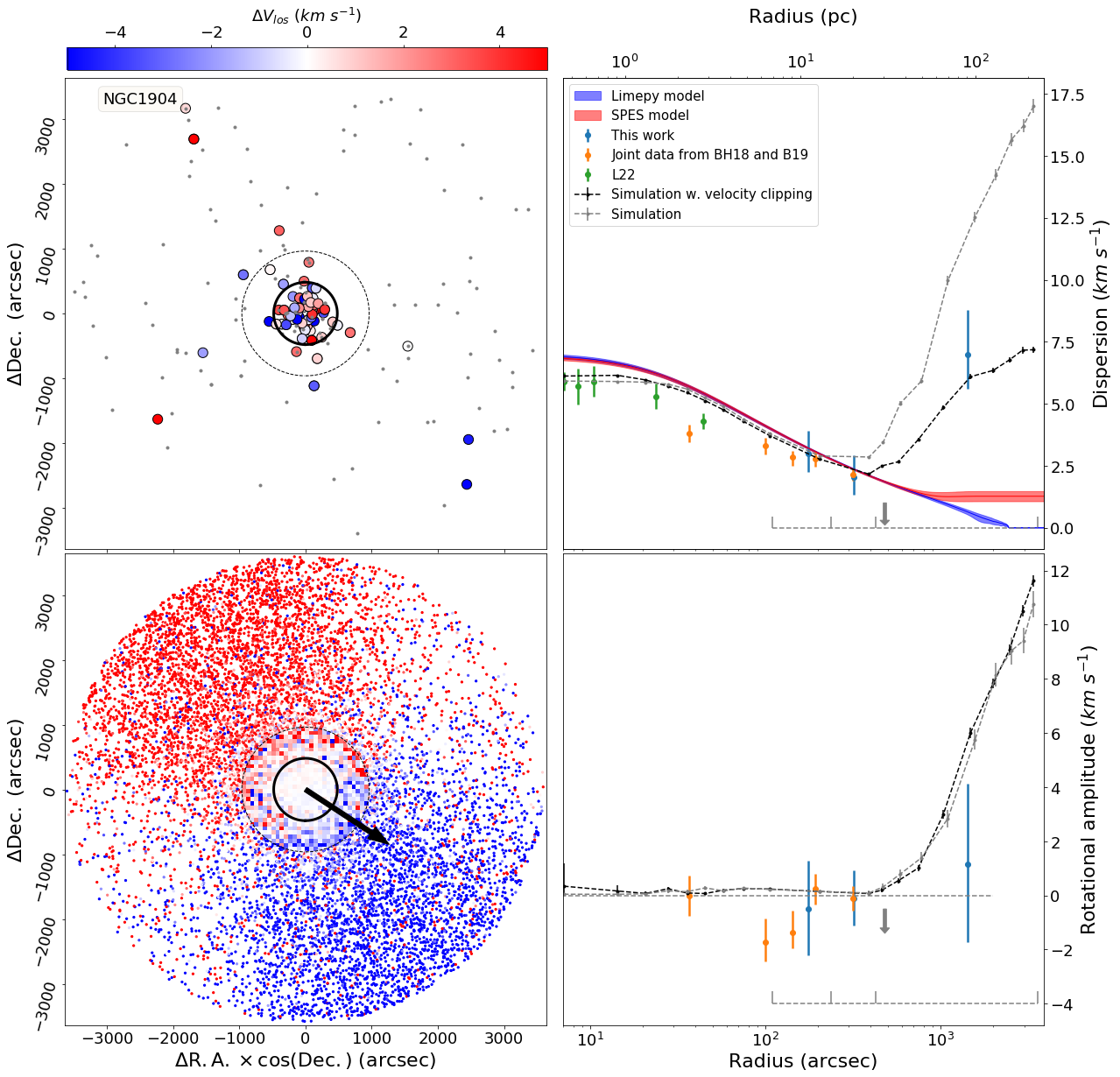}
    \caption{The same figure as Fig.~\ref{fig:N1261_results} for NGC~1904. Here the dashed circle has a radius of $2r_{\rm t}$. In the right panel, we include the data from \protect\citet[][coloured in green]{2022arXiv220307294L}, BH18/B19 (orange) to show the profile in the inner part.
    }
    \label{fig:N1904_results}
\end{figure*}

NGC~1904 has a present-day mass of $1.69\times10^{5}\ \mathrm{M_{\odot}}$, with a heliocentric distance of $\sim 13.1\ \mathrm{kpc}$, and a galactocentric distance of $\sim 19.1\ \mathrm{kpc}$. The orbit of this cluster suggests it has a small peri-centre with a galactocentric radius of $0.33\pm0.16  \, \mathrm{kpc}$, where the tidal effect from the MW on the cluster is significant. Efforts have been made to search for a signature of tidal interaction in the outskirts of this cluster: recently, \citet{2018MNRAS.474..683C} found some minor extra-tidal structures in the density map \citep[see also ][]{2018ApJ...862..114S,2022MNRAS.513.3136Z}, and a clear deviation from the {\sc limepy}/{\sc spes} models in the radial density profile of NGC~1904 around $\sim 300$ arcsec (note that the King truncation radius is $\sim 481.1$ arcsec), though no clear tidal feature that is directly associated with NGC~1904 has been found. 

We employed the two epochs available for NGC~1904, where we obtained 42 and 108 stars identified as {\it good\_star} from the first and second epochs correspondingly, with 42 stars in common. With the velocity clipping, we find 41 cluster members from epoch 1 and 58 members from epoch 2, with 40 members in common. We further exclude one star that has a measured metallicity $\mathrm{[Fe/H]} = -1.14 \pm 0.17$, and 11 stars brighter than $14$ mag. The metallicity of the members are shown in the third panel of Fig.~\ref{fig:Feh}; the mean is $\mathrm{[Fe/H]} = -1.66 \pm 0.01$ with an observed dispersion of $0.11 \pm 0.01$. The sky distribution of the members is shown in the top left panel of Fig.~\ref{fig:N1904_results}, where the members extend to $\sim 1$ degree away from the cluster centre. The simulation is shown in the corresponding bottom left panel, which has clear tidal features---many stars are scattered along with the tidal tails, while the random velocity is dominant within the truncation radius.

We calculated the dispersion and the rotation of NGC~1904 using Eq.~\ref{eq:model1}. For stars with two epochs of observation, only the spectrum with larger S/N was used in the calculations. The resultant profiles are shown in the right panels of Fig.~\ref{fig:N1904_results}, where the dispersion profile decreases within the truncation radius but increases in the last radial bin. The observations are in agreement with the results from BH18/B19 in the overlap region. For NGC~1904, we also include the data from \citet[][]{2022arXiv220307294L}, which extends the dispersion profile to the very inner part. 

The dispersion from the simulation and the {\sc limepy}/{\sc spes} models are also shown in the figure, which agree with each other well between $\sim 30 - 200$ arcsec. The models predict a larger dispersion within $30$ arcsec and a smaller dispersion beyond $200$ arcsec. Both the simulation and the models predict reasonable dispersion profiles compared to the observations around $\sim 100-300$ arcsec. In the inner part, though the {\sc limepy}/{\sc spes} models predict larger dispersion, the simulation agrees well with the observation from \citet[][]{2022arXiv220307294L}. There are some discrepancies around $30-50$ arcsec, where the simulation/models predict a larger dispersion compared to the data from BH18/B19. For the outskirts, the simulation also shows a significant increase in the dispersion profile, which corresponds to the tidal tails presented in the bottom left panel of Fig.~\ref{fig:N1904_results}. The velocity clipping excludes stars with large relative velocities, leading to a better fitting dispersion profiles. The observed dispersion in the last radial bin is $6.99^{+1.80}_{-1.37}\ \mathrm{km\ s^{-1}}$, and the nearest radial bin in the simulation with velocity clipping has a dispersion of $6.09 \pm 0.11\ \mathrm{km\ s^{-1}}$, in agreement with the observation.

We cannot confirm any rotation signal from our observation. In the outermost bin, the rotation amplitude is $1.14^{+2.97}_{-2.88}\ \mathrm{km\ s^{-1}}$, which is consistent with zero. In the inner part around $100-200$ arcsec ($\sim 10\ \mathrm{pc}$), the data from BH18/B19 show some rotation with amplitude of $-1.74^{+0.88}_{-0.69}\ \mathrm{km\ s^{-1}}$ (the amplitude in the second radial bin in BH18/B19, the negative value in the figure is due to the direction with respect to the position angle). However, with the large uncertainty, our data are not able to confirm the internal rotation.  

\subsection*{NGC~1851}

\begin{figure*}
    \centering
    \includegraphics[width=2\columnwidth]{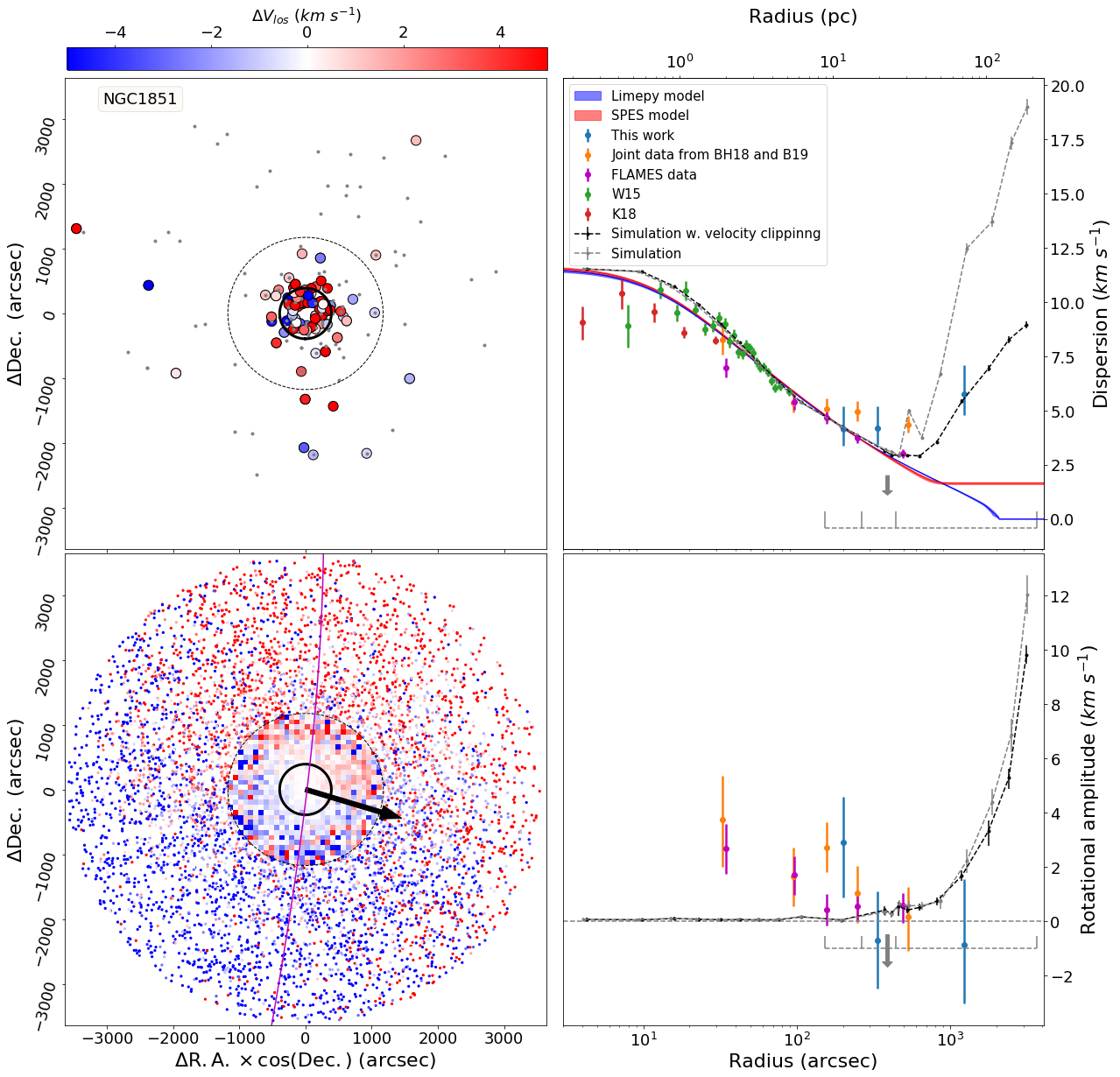}
    \caption{The same figure as Fig.~\ref{fig:N1261_results} for NGC~1851. Here the stream data are from \protect\citet[][]{2021ApJ...914..123I}, and the dashed circle has a radius of $3r_{\rm t}$. In the right panel, we include the data from the joint data from BH18/B19 (the orange errorbars), the data from FLAMES observation (the magenta errorbars), \protect\citet[][W15]{2015ApJ...803...29W} (the green error bars), and \protect\citet[][K18]{2018MNRAS.473.5591K} (the red error bars).
    }
    \label{fig:N1851_results}
\end{figure*}

NGC~1851 is the most massive cluster among our targets, with a present-day mass of $3.02\times10^{5}\ \mathrm{M_{\odot}}$ \citep[][]{2017MNRAS.471.3668S}. It is located at a heliocentric distance of $\sim 11.9\ \mathrm{kpc}$ and a galactocentric distance of $\sim 16.7\ \mathrm{kpc}$. Similar to NGC~1261, NGC~1851 is known to be embedded in a substantial stellar envelope \citep{2009AJ....138.1570O,2014MNRAS.445.2971C,2018MNRAS.473.2881K}. \citet{2018MNRAS.474..683C} found a clear break from a power-law in the radial density profile at $\sim 10$ arcmin, with a clear extension of stars out to $\sim 80$ arcmin in the density map. The spectroscopic study by \citet{2012MNRAS.426.1137S} found cluster members out to $\sim 30$ arcmin using the radial velocity, and candidate members of the cluster have been found to a larger distance \citep[e.g.,][]{2009A&A...503..755M,2009ApJ...697L..22Z,2014MNRAS.442.3044M,2014A&A...572A..30K}. 

The observations of NGC~1851 were obtained in two epochs as shown in Tab.~\ref{tab:observation_log}, where 81 and 59 stars are classified as {\it good\_star} in the first and second epochs correspondingly. With the velocity clipping Eq.~\ref{eq_selection}, we have 70 members from the first epoch and 50 members from the second epoch, among which we have 50 stars with two observations. The metallicities of the selected members are shown in the bottom panel of Fig.~\ref{fig:Feh}, from which we found a mean metallicity of $\mathrm{[Fe/H]} = -1.19 \pm 0.02$ with a dispersion of $0.14\pm0.02$ in good accord with the value, $\mathrm{[Fe/H]} = -1.18$, given in the \citet{Harris10} on-line catalogue. We note again that the observed dispersion could be dominated by the metallicity measurement errors. The sky distribution of the members is shown in the top left panel of Fig.~\ref{fig:N1851_results}, where it is evident that there are many members scattered beyond the truncation radius of the King model. The corresponding simulation is shown in the bottom left panel, where the data within $3 \, r_{\rm t}$ are presented as a heat map. 

We estimated the dispersion and the rotation of NGC~1851 by fitting Eq.~\ref{eq:model1} with the data, and the results are shown in the right panels of Fig.~\ref{fig:N1851_results}. We have also incorporated the data from \citet[][]{2015ApJ...803...29W}, \citet{2018MNRAS.473.5591K}, BH18/B19, and the observational data from FLAMES to extend profiles to the inner parts of the cluster, which match well with each other in the overlap regions of the data sets. In the outer part beyond $\sim 200$ arcsec/$10$ pc, the observed dispersion remains relatively flat at $\sim 4.6\ \mathrm{km\ s^{-1}}$ (the average from the first two bins of our results and the last three bins of BH18/B19) with increasing radius, which agrees well with the dispersion profile derived by \citet{2011A&A...525A.148S}. However, the outermost bin does show a larger dispersion with an amplitude of $5.76^{+1.33}_{-0.97}\ \mathrm{km\ s^{-1}}$. The {\sc limepy}/{\sc spes} models are also included, together with the simulation, as a comparison to the data. Within the truncation radius, the {\sc limepy}/{\sc spes} models and the simulation have very similar predictions for the dispersion profile, while in the inner part the data present to have lower velocity dispersion. Unlike other GCs in this work, the dispersion profile of NGC~1851 deviates from the {\sc limepy}/{\sc spes} models within the truncation radius, starting within the first radial bin based on our observations and BH18/B19, where the dispersion data do not decrease as expected by the {\sc limepy}/{\sc spes} models. This deviation might due to the intense interaction with the MW. Outside the truncation radius, the simulation shows a sharp rise in the dispersion, where the observed dispersion is well reproduced by the simulation with velocity clipping. The last radial bin has a dispersion of $5.77^{+1.33}_{-0.97}\ \mathrm{km\ s^{-1}}$, where the nearest simulation with velocity clipping has a dispersion of $5.43\pm0.11\ \mathrm{km\ s^{-1}}$.

As for rotation, based on our observations and BH18/B19, there is a clear rotational signal of $\sim 3\ \mathrm{km\ s^{-1}}$ within the truncation radius. Though the FLAMES data show signals consistent with zero around $100-300$ arcsec, they agree well with BH18/B19 in the inner part with a clear rotation signal. In addition, there does not seem to be strong evidence for rotation at and beyond the truncation radius, where the results are consistent with zero within the uncertainties. However, the simulation does not show evidence of rotation within the truncation radius, but has a large increasing rotation signal beyond $\sim 1000$ arcsec. Here we compare our results with the previously detected rotation signal. The major rotation signal is in the first radial bin ($153 < r < 266$ arcsec). The amplitude of this signal is $2.90^{+1.68}_{-2.02}\ \mathrm{km\ s^{-1}}$, with a position angle of $52^{\circ}.8^{+26.4}_{-49.8}$. The rotation signal of NGC~1851 was also detected in several previous studies in the literature. \citet[][]{2015A&A...573A.115L} found a rotation of $1.65\ \mathrm{km\ s^{-1}}$, with the position angle of $50^{\circ}$. This result is in good agreement with our measurement. We also note that the smaller amplitude might be because they use a larger data bin that includes members within about $600$ arcsec, where our results show that the signal tends to be small. \citet[][]{2018MNRAS.473.5591K} also found the rotation signal of $\sim 2\ \mathrm{km\ s^{-1}}$ over a range of radii, but their detection is concentrated within the central region of this cluster ($r < 100$ arcsec) \citep[see also][]{2012A&A...538A..18B}.

\section{Discussion}
\label{sec:conclusion}

In this paper we have presented the chemo-dynamical results of four GCs---NGC1261, NGC4590, NGC1904, NGC1851---out to 1 degree from the cluster centres, with the AAT + 2dF/AAOmega and VLT-FLAMES, extracting the metallicity and velocity information of each target. We find the members, selected with the velocity clipping, are distributed beyond the truncation radius of the King models in all four clusters. $N$-body simulations for each cluster, as well as the theoretical predictions from the {\sc limepy}/{\sc spes} models, are used as comparisons to interpret the observation.

We applied the same velocity clipping Eq.~\ref{eq_selection} to both the data and the simulations. In the simulations, this clipping excludes small amount of stars that have very large relative velocities---in the outermost 9 radial bins in each simulation, we exclude 24 from 3621 stars in NGC~1261; 0 from 38805 stars for NGC~4590; 752 from 17094 stars in NGC~1904, and 475 from 15807 stars for NGC~1851. Excluding those stars primarily influence the dispersion and barely change the rotation profiles as the Fig.~\ref{fig:N1261_results}-\ref{fig:N1851_results} show. Given the rotation amplitude and dispersion in the outskirts of the observations, the percentage of potential members that might be excluded by the velocity clipping should be smaller than the simulation, especially for NGC~1904 and NGC~1851, whose simulations show very large dispersion and rotation in the outskirts.

In the simulation of NGC~1261, we find a clear increasing dispersion, and an increasing rotation signal beyond the truncation radius, which comes from the escaped stars around the cluster. The observation supports the results from the simulation with a large dispersion in the last radial bin, while the simulation with velocity clipping predicts lower dispersion. In the observations, we found the outermost three stars contribute most to the dispersion, with the LOS velocity of $(10.87\pm0.57, 7.51\pm1.10, 11.54\pm0.55)\ \mathrm{km\ s^{-1}}$, and the corresponding metallicity of $(-1.65\pm0.06, -1.32\pm0.08, -1.33\pm0.08)\ dex$. The first star deviates $\sim 0.3$ dex from the mean value, suggesting that this star is unlikely to be the member of NGC~1261. We estimated that the dispersion will be $4.14^{+1.60}_{-1.06}\ \mathrm{km\ s^{-1}}$, and this is only marginally inconsistent with the simulation ($2.87\pm0.08\ \mathrm{km\ s^{-1}}$). The metallicities of the last two stars suggest that they are very likely to be the cluster members. Their positive LOS velocities, as shown in the top left panel of Fig.~\ref{fig:N1261_results}, also agree well with the simulation, suggesting that they are the cluster members that are either in the process of leaving or already in the tail. 

While the corresponding simulation suggests a much larger dispersion in the outskirts, a similar increasing dispersion is also found in NGC~1904. In the last radial bin, two stars contribute significantly to the large dispersion, with LOS velocities of $(-18.80\pm0.95, 13.55\pm2.38)\ \mathrm{km\ s^{-1}}$, and corresponding metallicities of $(-1.84\pm0.07, -1.38\pm0.15)\ dex$. We cannot confirm the membership of the two stars based on the metallicity. If the two stars are excluded from the sample, the dispersion would be $2.45^{+1.08}_{-0.81}\ \mathrm{km\ s^{-1}}$, which is roughly the same scale as the dispersion in the second radial bin, resulting in a flat dispersion profile in agreement with the estimation from \citet{2011A&A...525A.148S}.

Our final member distribution of NGC~4590 does not effectively extend beyond the truncation radius. Our results agree well with the {\sc limepy}/{\sc spes} model, and only marginally inconsistent with the simulation, where the simulation predicts a smaller dispersion profile beyond $\sim 200$ arcsec. 

Similar to NGC~1261 and NGC~1904, the dispersion profile of NGC~1851 also deviates from {\sc limepy}/{\sc spes} model beyond the truncation radius, while it becomes flat outwards in agreement with \citet{2011A&A...525A.148S}. Interestingly in this cluster, unlike the other GCs, the deviation from the models occurs likely within the truncation radius---the flat profile starts from the radius of $\sim 3-4$ arcmin. However, the power-law break of the NGC~1851 density profile is around $\sim 10$ arcmin, which is much larger than the radius where the flattening begins. Hence it is hard to explain the deviation with the stellar envelope alone. Dynamical models suggests that the presence of dark matter naturally leads to an inflated/raising dispersion profile \citep[e.g.,][]{2015ApJ...808L..35T,2017MNRAS.471L..31P}, however, no direct evidence points to the existence of dark matter in GCs.  Besides, the dynamical model in \citet[][]{2017MNRAS.471L..31P} is based on the assumption of equilibrium, which cannot be applied to clusters that are losing stars due to tidal interaction (the same issue also applies to \citet[][]{2021ApJ...922..104C}). Our simulation shows that the tidal interaction can produce an increasing dispersion profile beyond the truncation radius without dark matter---the simulations show larger dispersion in the outskirts of the cluster, and applying the same velocity clipping results in dispersion profiles that match the simulation well in NGC~1851 and NGC~1904, and is consistent with NGC~1261.

For NGC~1261, NGC~1851 and NGC~4590, we detected a non-zero rotation signal in several radial bins. The internal rotation signal of NGC~1261 is likely to be the bias from the binned data---the dispersion of the LOS velocity is interpreted to be rotation by the fitting. The rotation in the outskirts of NGC~1261 has a position angle of $87^{\circ}.1^{+48.1}_{-48.7}$, which is consistent with the simulation ($111^{\circ}.0^{+19.0}_{-12.6}$). We calculated the orbital angular momentum vector of NGC~1261 within the MW potential, which has a position angle of $116^{\circ}.9$ and is consistent with the rotation signal in the simulation and the observation. The rotation axis of potential escapers is expected to be aligned with the orbital angular momentum, and hence suggests that the observed members in the outskirts of NGC~1261 are unbound to the cluster. NGC~1851 has internal rotation within the truncation radius, where the simulation is dominated by random motion. This signal agrees well with the data from BH18/B19 and existing measurements from literature (see Sec.~\ref{sec:results}), supporting that NGC~1851 has internal rotation within the truncation radius. This rotation signal can be the remnant of the primordial regular motion, which is gradually erased by the internal/external dynamical processes \citep[see e.g., ][ and references therein]{2019MNRAS.485.1460S}, and could be responsible for the misalignment of the stream track in our simulation and the observations in Fig.~\ref{fig:sim_stream} \citep{2010MNRAS.408L..26P}. We note that based on our results, the rotation of GCs could be differential where the amplitude and position angle could be different at different radii \citep[see also ][]{2018MNRAS.473.5591K}.

In NGC~4590, we see only one detection of non-zero rotation signal with large uncertainty, and the rotation profile in NGC~1904 is consistent with zero. Though we detect rotation with similar amplitude from BH18/B19 in NGC~1904, the large uncertainty in both NGC~4590 and NGC~1904 suggests that whether the signal comes from regular or random motion is still uncertain, and it requires further observation to confirm and distinguish the signal from noise.

In the simulation, we see that both the dispersion profiles and the rotation profiles increase significantly beyond the truncation radius due to the escaped stars in the tidal arms. The observations also present the same feature, i.e. that the dispersion increases in the outskirts. However, we do not observe the same rotation in NGC~1904 and NGC~1851 expected by the simulation. One reason might be the low number of members, and the large dispersion in the outskirts of NGC~1904 and NGC~1851. We found the rotation profiles in the last radial bin of both NGC~1904 and NGC~1851 possess large uncertainties, suggesting that there might be rotation signal that is indistinguishable with the large velocity dispersion based on our data. A larger sample set that could possibly reduce the uncertainties is required to resolve the signal in the future.

In NGC~3201 \citep[from our previous work,][]{2021MNRAS.502.4513W}, we detected an increasing dispersion in the outskirts of the cluster larger than the {\sc limepy}/{\sc spes} models prediction. An $N$-body simulation applied to NGC~3201 that matches the inner part of the cluster shows a $2\sigma$ lower dispersion in the outskirts. We explored the factor that can cause the large velocity dispersion, including binaries and black hole dynamics, but none of them alone can fully explain the dispersion. In some clusters,it has been suggested that this large dispersion is a sign of the presence of the dark matter content \citep[e.g.,][]{2021ApJ...922..104C}. In this work, we also detect flat/increasing dispersion profiles larger than the {\sc limepy}/{\sc spes} models prediction. However, the increasing dispersion profiles are reproduced in, and can be explained by, the $N$-body simulations by the interaction with the MW without dark matter content within the clusters. In addition, \citet[][]{2022MNRAS.513.3136Z} found that all the clusters known to possess significant envelopes have also recently been found from Gaia to possess tidal tails, suggesting that the envelopes could be consistent with tidal erosion of clusters, as an alternative explanation to the dark matter possibility. Hence, probing the presence of dark matter within the cluster requires detailed modelling to include/exclude different factors. 

\section*{Acknowledgements}
ZW acknowledge the support of the National Natural Science Foundation of China (NSFC, grant No. 12173037), the China Manned Space Project with NO. CMS-CSST-2021-A04 and Cyrus Chun Ying Tang Foundations.
WHO gratefully acknowledges financial support through the Paulette Isabel Jones PhD Completion Scholarship at the University of Sydney.  MG acknowledges support from the Ministry of Science and Innovation (EUR2020-112157, PID2021-125485NB-C22) and from Grant CEX2019-000918-M funded by MCIN/AEI/10.13039/501100011033. VHB acknowledges the support of the Natural Sciences and Engineering Research Council of Canada (NSERC) through grant RGPIN-2020-05990. EB acknowledges financial support from a Vici grant from the Netherlands Organisation for Scientific Research (NWO). GFL received no funding to support this research.

Based in part on data acquired through the Australian Astronomical Observatory. We acknowledge the traditional owners of the land on which the AAT stands, the Gamilaraay people, and pay our respects to elders past, present and emerging.

Based on observations collected at the European Southern Observatory under ESO programme(s) 0102.D-0164A and/or data obtained from the ESO Science Archive Facility with DOI(s) under \href{https://doi.org/10.18727/archive/27}{https://doi.org/10.18727/archive/27}.

Parts of this work were performed on the OzSTAR national facility at Swinburne University of Technology. The OzSTAR program receives funding in part from the National Collaborative Research Infrastructure Strategy (NCRIS) Astronomy allocation provided by the Australian Government.

This research has been partly funded through the INAF Main Stream grant 1.05.01.86.22 "Chemo-dynamics of globular clusters: the Gaia revolution" (PI: Pancino). 

\section*{Data Availability}
The data underlying this article may be made available on reasonable request to the corresponding author.

Our NBODY6+P3T code is free for use and is available at https://github.com/anthony-arnold/nbody6-p3t/tree/v1.1.0.




\bibliographystyle{mnras}
\bibliography{ref}


\onecolumn
\appendix

\section{Comparisons to streams}
In this appendix, we present the stellar distribution in the simulations and compare to the observed stellar streams that are considered to be associated with the GCs mentioned in this paper. Here the stream tracks are calculated by \citet[][]{2022arXiv220410326M} with polynomial fitting, where the NGC~1261 data come from \citet[][]{2021ApJ...914..123I}, the NGC~1851 data come from \citet[][]{2021ApJ...914..123I} as well, and the NGC~4590 data come from \citet[][]{2019MNRAS.488.1535P}. We notice that the simulations match the NGC~1261 and NGC~4590 tidal tails well, especially NGC~4590, where the observed stream is well represented by the simulation across a large angular extent. The track of NGC~1851 streams deviates strongly from the ridge of the stellar distribution in the simulation. We note that the internal rotation in the progenitor system could be responsible for the mismatch if the rotating axis is misaligned with the orbital angular momentum \citep[e.g., Fig,1 in][]{2010MNRAS.408L..26P}.

\begin{figure*}
    \centering
    \includegraphics[width=\textwidth]{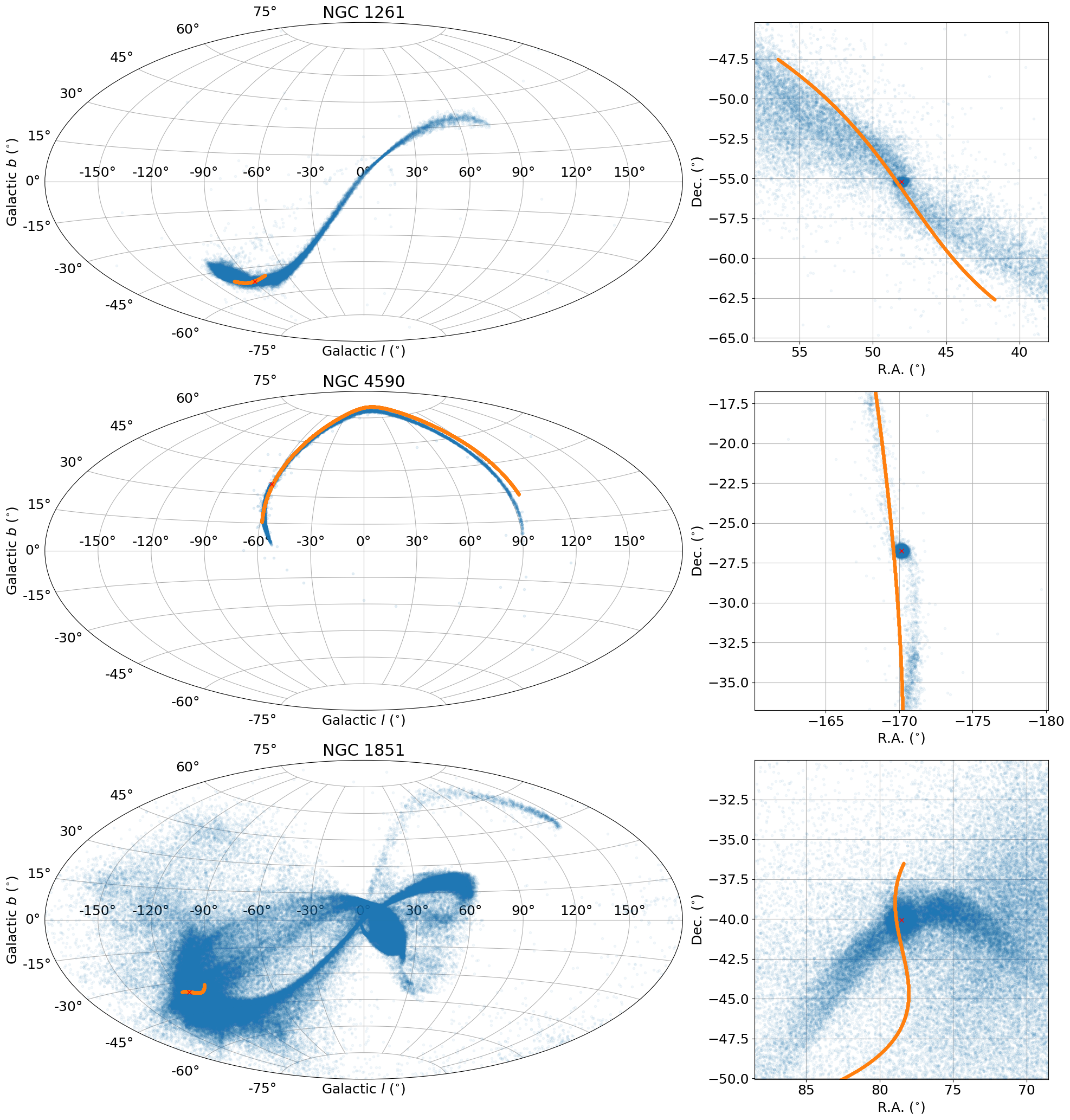}
    \caption{The comparison between the simulations and the corresponding observed streams in the Galactic coordinate. The blue points are stars in the simulation, the orange tracks are the stream tracks and the red crosses are the location of the clusters. From the top the bottom are NGC~1261, NGC~4590 and NGC~1851 respectively, while the right panel shows the zoom-in view around the corresponding clusters in Equatorial coordinate. }
    \label{fig:sim_stream}
\end{figure*}

\bsp	
\label{lastpage}
\end{document}